\documentclass[reprint,prl,amsmath,amssymb,aps,superscriptaddress,nofootinbib]{revtex4-2}

\usepackage{physics2,derivative,cancel,color,amsthm,graphicx}
\usepackage[mathlines]{lineno}
\usepackage{fontawesome5}
\usepackage{hyperref}
\usephysicsmodule{ab,ab.braket}

\DeclareMathOperator{\jsd}{sd}

\begin{document}

\title{Stable Evaluation of Lefschetz Thimble Intersection Numbers:\\Towards Real-Time Path Integrals}

\author{Yutaro Shoji}
\email{yutaro.shoji@ijs.si}
\affiliation{Jo\v{z}ef Stefan Institute, Jamova 39, 1000 Ljubljana, Slovenia}
\affiliation{Centre For Cosmology and Science Popularization (CCSP), SGT University, Gurugram,
Delhi-NCR, Haryana 122505, India}

\author{Katarina Trailović}
\email{katarina.trailovic@ijs.si}
\affiliation{Jo\v{z}ef Stefan Institute, Jamova 39, 1000 Ljubljana, Slovenia}
\affiliation{Faculty of Mathematics and Physics, University of Ljubljana, Jadranska 19, 1000 Ljubljana, Slovenia}

\date{\today}

\begin{abstract}
    We introduce a robust numerical method for determining intersection numbers of Lefschetz thimbles in multivariable settings. Our approach employs the multiple shooting method to solve the upward flow equations from the saddle points to the original integration cycle, which also enables us to determine the signs of the intersection numbers. The method demonstrates stable and reliable performance, and has been tested for systems with up to $20$ variables, which can be further extended by adopting quadruple-precision arithmetic. We determine intersection numbers for several complex saddle points in a discretized path integral, providing new insights into the structure of real-time path integrals. The proposed method is broadly applicable to a wide range of problems involving oscillatory integrals in physics and mathematics. \href{https://github.com/KTrailovic/Lefschetz_Thimble_Intersection_Numbers}{\faGithub}
\end{abstract}

\maketitle

\section{Introduction}
Oscillatory integrals present significant challenges in numerical simulations due to the so-called sign problem, which undermines the reliability of conventional integration techniques. Picard-Lefschetz theory offers a mathematically rigorous framework for analyzing such integrals, particularly in multivariable cases \cite{Pham1983,Kaminski1994,Howls1997}.
In recent years, this approach has found applications to real-time Feynman path integrals \cite{Witten:2010cx,Witten:2010zr}, finite-density quantum chromodynamics (QCD) \cite{Cristoforetti:2012su,Cristoforetti:2013wha,Aarts:2013fpa,Fujii:2013sra}, quantum tunneling \cite{Tanizaki:2014xba,Cherman:2014sba,Ai:2019fri,Hayashi:2021kro,Nishimura:2023dky,Garbrecht:2024end,Garbrecht:2025tos}, gravitational lensing \cite{Feldbrugge:2019fjs,Shi:2024msk}, quantum cosmology and gravity \cite{Feldbrugge:2017kzv,Feldbrugge:2017fcc,DiazDorronsoro:2017hti,Jia:2022nda,Honda:2024aro,Chou:2024sgk,Chen:2024qmn,Ailiga:2025fny,Ailiga:2025osa}, and solid-state physics \cite{Mukherjee:2014hsa}.

The Picard-Lefschetz theory provides a framework for decomposing the real integration domain $\mathcal Y=\mathbb R^L\subset\mathbb C^L$ into a sum over steepest-descent cycles in $\mathbb{C}^L$, known as Lefschetz thimbles. Each thimble is associated with a complex saddle point, yielding the decomposition
\begin{equation}
    \int e^{\frac{\mathcal I(x)}{\hbar}}\,\odif[order=L]{x}
    = \sum_\sigma n_\sigma \int_{\mathcal J_\sigma} e^{\frac{\mathcal I(z)}{\hbar}}\,\odif[order=L]{z}.
\end{equation}
Here, $\mathcal I(z)$ is a holomorphic function and $\hbar$ is a constant. We use the notation $x=(x_0~\ldots~x_{L-1})$ for real variables and $z=(z_0~\ldots~z_{L-1})$ for complex variables.
A Lefschetz thimble is denoted by $\mathcal J_\sigma$, and is defined as the set of all points in $\mathbb{C}^L$ that can be reached by integrating the downward flow equation $\partial z_i/\partial u = -\overline{\partial \mathcal I/\partial z_i}$ from the saddle point $z(u=-\infty) = z_\sigma$. Notice that there are $L$ independent solutions around a saddle point and thus $\mathcal J_\sigma$ is an $L$-dimensional submanifold. Along each thimble, the imaginary part of the exponent $\mathcal I$ remains constant while the real part decreases monotonically, ensuring that the integrals over $\mathcal J_\sigma$ are convergent and well-defined.

The coefficient $n_\sigma$ is an integer that specifies the contribution of each thimble to the integral over the original cycle $\mathcal Y$. It is given by the intersection number between $\mathcal Y$ and the upward flow cycle $\mathcal K_\sigma$ in the sense of homology, $n_\sigma = \langle \mathcal Y, \mathcal K_\sigma \rangle$. Since both $\mathcal Y$ and $\mathcal K_\sigma$ are $L$-dimensional submanifolds embedded in $2L$-dimensional space, their intersection generically consists of isolated points. 
While our method is capable of identifying multiple intersection points, in this work we focus on the generic case where $n_\sigma = 0$ or $\pm 1$. Cases with $|n_\sigma| > 1$ typically occur only in the presence of additional symmetries or degeneracies, which are beyond the scope of the present analysis.

Determining intersection numbers in multivariable integrals has long been a challenging and unresolved problem. One major difficulty is that the upward flow equations often exhibit chaotic behavior over extended flow times, resulting in extreme sensitivity to initial conditions and numerical errors. This makes the single shooting method largely impractical. Furthermore, the computational difficulty increases exponentially with the number of variables, which has limited previous studies mostly to cases involving only one or two variables.

For small or effectively small number of variables, several methodologies have been proposed for determining intersection numbers. Feldbrugge et al.~\cite{Feldbrugge:2019fjs} demonstrated a smooth deformation of the original integration cycle $\mathcal Y$ into a sum over Lefschetz thimbles, enabling explicit decomposition and direct computation of intersection numbers. Fujimori et al.~\cite{Fujimori:2022lng} constructed exact upward flows by exploiting symmetries. A matrix formalism has also been proposed for this purpose~\cite{shanin2025}. Indirect approaches include inference via the Stokes phenomenon~\cite{Kanazawa:2014qma,Fujii:2015bua} and comparison of different integration contours~\cite{Lawrence:2023woz}. Additionally, connections to the Maslov index have been explored~\cite{Sueishi:2020rug}.

In this work, we present a robust and efficient numerical method for determining upward flows that connect saddle points $z_\sigma$ to the original integration cycle $\mathcal Y$ in multivariable settings. Our approach leverages the multiple shooting technique \cite{10.1145/355580.369128,keller1976numerical,Biegler2010}, widely employed in fields such as celestial mechanics and chemical engineering to address nonlinear systems exhibiting strong sensitivity to initial conditions. The algorithm achieves high reliability and computational efficiency, enabling the identification of relevant flows even in systems with tens of variables within a minute on a single computational thread. In addition, the method stably propagates a tangent space of $\mathcal K_\sigma$ from the saddle point to the intersection point, allowing us to determine also the sign of the intersection number.

We apply the method to compute intersection numbers in a discretized path integral for a quantum mechanical system, which is particularly important for the understanding of real-time path integrals. Within the limitations of the discretized approximation, our approach enables the explicit determination of intersection numbers for nontrivial complex saddle points, addressing an open problem in the field.

\section{Multiple Shooting Method}
A standard technique for solving boundary value problems is the single shooting method. In this approach, solutions are constructed by integrating the flow equations forward from initial conditions $z(0) = z_\sigma + \epsilon$, where $\epsilon$ is a small perturbation. The desired solution is obtained by adjusting $\epsilon$ to satisfy the boundary conditions at the final point. However, in systems characterized by strong sensitivity to initial conditions, even tiny variations in $\epsilon$ can result in exponentially diverging trajectories as the flow traverses regions of instability. This exponential sensitivity severely limits the practical applicability of the single shooting method for such problems.

To address this issue, we employ the multiple shooting method \cite{10.1145/355580.369128,keller1976numerical,Biegler2010}. This approach partitions the integration domain into several subintervals, within which the dependence on their own initial conditions remains approximately linear. Solutions are constructed independently on each subinterval, and continuity is enforced by matching the endpoints of adjacent segments, together with the original boundary conditions. The resulting system is formulated as a nonlinear optimization problem, which can be efficiently solved using algorithms such as Newton's method. This overcomes the exponential sensitivity to initial conditions, since perturbations are restricted to propagate only linearly during optimization. This stabilization significantly improves the robustness and convergence properties of the algorithm, which has been extensively validated in various fields. Moreover, its stability and convergence have been well understood mathematically.

In the following, we present our formulation.
The upward flow equation, $\partial z_i/\partial u = \overline{\partial \mathcal I/\partial z_i}$, often exhibits rapid variations in the flow velocity, which can hinder numerical stability. Therefore, we use a normalized form of the upward flow given by $\odif{z_i}/\odif{s}=\overline{\partial \mathcal I/\partial z_i}/|\partial \mathcal I/\partial z_i|$.

To describe the boundary conditions, it is more useful to decompose the complex variables into their real parts and imaginary parts as
$Z=(\Re(z)~\Im(z))$.
Around a saddle point $Z=Z_\sigma$, we compute the eigenvalues $\pm\lambda_i$ and the corresponding eigenvectors $W_i^\pm$ of the Hessian matrix $H=\sigma_3\otimes\Re{\rm d}^2\mathcal I/{\rm d}z^2-\sigma_1\otimes\Im{\rm d}^2\mathcal I/{\rm d}z^2$, where $\sigma_a$'s are Pauli matrices, $\lambda_i>0$ and $W_i^\pm\cdot W_i^\pm=1$.
Notice that there are the same number of positive and negative eigenvalues because the Hessian satisfies $\epsilon^\dagger H \epsilon=-H$ with $\epsilon=i\sigma_2\otimes1_{L\times L}$. We assume that there is no degenerate saddle point, and hence there is no zero eigenvalue.
This defines two $2L\times L$ matrices, $W^\pm=(W_0^\pm~\cdots~W_{L-1}^\pm)$.

The boundary conditions consist of three parts:
the first condition, $|Z(0)-Z_\sigma|-\delta r=0$, fixes the shifting freedom along the solution (anchor condition), the second condition, $(W^-)^t(Z(0)-Z_\sigma)=0$, selects the upward flow around the saddle point, and the last condition, $(0_{L\times L}~1_{L\times L})Z(s_f)=0$, sets imaginary parts to zero at $s=s_f$. Here, $\cdot^t$ denotes transpose, $\delta r>0$ is a small parameter and $s_f>0$ is the final time.

The multiple-shooting method solves the upward flow equation by partitioning the integration interval into $N-1$ subintervals of length $\delta s$. Within each subinterval, the flow is precisely integrated, and the continuity condition is imposed.
In total, the system comprises $2L+1$ boundary conditions and $2L(N-1)$ continuity conditions, corresponding to $2LN+1$ equations for the $2LN$ variables $Z^{(k)}$ and the step size $\delta s$. The resulting nonlinear system is efficiently solved using Newton's method. Notably, by treating $\delta s$ as an optimization variable with fixed $N$, the final integration time $s_f=(N-1) \delta s$ is determined self-consistently as part of the solution.

A key feature of Newton's method in this context is its clear convergence behavior: when a solution exists, the optimization sequence converges rapidly, typically within 100 iterations, until limited by numerical precision. Conversely, in the absence of a solution, the sequence either oscillates or diverges. This characteristic provides a practical diagnostic for determining the existence of intersections.
The Newton's method also propagates the tangent space of $\mathcal K_\sigma$ at the saddle point to that at the intersection point, which determines the sign of $n_\sigma$ for given orientation of $\mathcal J_\sigma$.

A detailed exposition of the algorithm is provided in Supplemental Material.
We provide an implementation of our algorithm in Mathematica, available at \cite{mathematica_github}. A Rust implementation is also available upon request.
\section{Results}
\subsection{Three-variable Airy-type integral}
To illustrate the robustness of our method, we first consider a tractable example involving three variables. Specifically, we examine the exponent
\begin{align}
    \mathcal{I}(x) &=i\left[\frac{x_0^3+x_1^3+x_2^3}{3}-x_0x_1-x_1x_2-x_2x_0\right.\nonumber\\
    &\hspace{5ex}\left.+c_0x_0+c_1x_1+c_2x_2\right],
\end{align}
where the $c_i$'s are complex parameters.
This resembles the exponent of the Airy integral for each variable, and the quadratic terms induce mixing which make the integral non-trivial.
According to B\'ezout's theorem, this system admits eight saddle points on $\mathbb C^3$.

This oscillatory integral is convergent by slightly deforming $\mathcal Y$ such that the integrand dumps exponentially.
This allows for a direct comparison between results obtained via numerical integration and those via the saddle-point approximation, serving as a benchmark to validate the accuracy and reliability of our approach.

To reduce the number of free parameters, we set $c_n = 0.5\,e^{i(n+1)\alpha}$ with $0 \leq \alpha < 2\pi$. The parameter $\hbar$ must be sufficiently small to ensure the validity of the saddle-point approximation. However, taking too small values of $\hbar$ makes direct numerical integration imprecise due to increased oscillatory behavior. In the following analysis, we choose $\hbar = 0.05$, which yields an expected error of approximately $5\%$ in the saddle-point approximation. As for the parameter in the anchor condition, we take $\delta r=0.01$ for most of $\alpha$ and relax it up to $\delta r=0.03$ when we encounter a numerical difficulty. We set $N=200$.

The saddle-point approximation for the integral is given by $\int e^{\frac{\mathcal I(x)}{\hbar}}\,\odif[order=L]{x}= \sum_{\sigma}n_\sigma A_\sigma e^{\frac{\mathcal I(z_{\sigma})}{\hbar}}[1+\mathcal O(\hbar)]$, where $A_\sigma=\det J\prod_i(2\pi\hbar/\lambda_i)^{1/2}$ with $J$ denoting the Jacobian that maps the local real coordinates on $\mathcal J_\sigma$ to the complex variables of the embedding. Each column of $J$ represents the pushforward of a coordinate vector in the parameter space and thus corresponds to a tangent vector in the ambient space. At the saddle, the tangent space of the thimble is spanned by the eigenvectors $W^-$, allowing us to choose coordinates such that $J_{ab} = W^-_{ab} + i W^-_{(L+a)b}$.

\begin{figure}[t]
    \centering
    \includegraphics[width=0.234\textwidth]{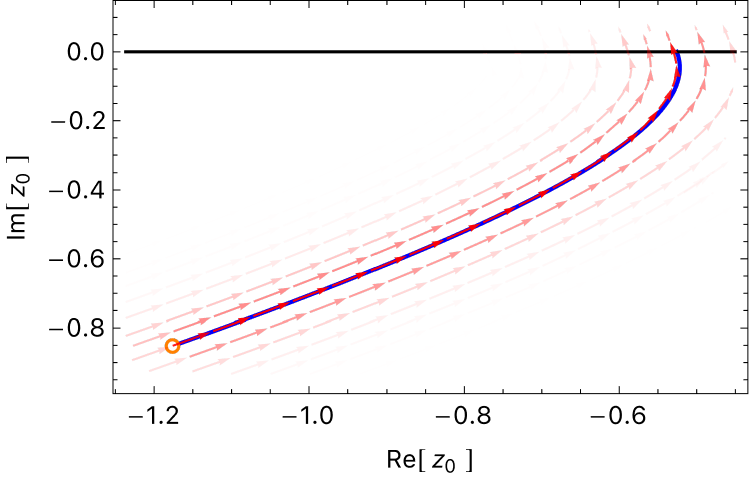}
    \includegraphics[width=0.234\textwidth]{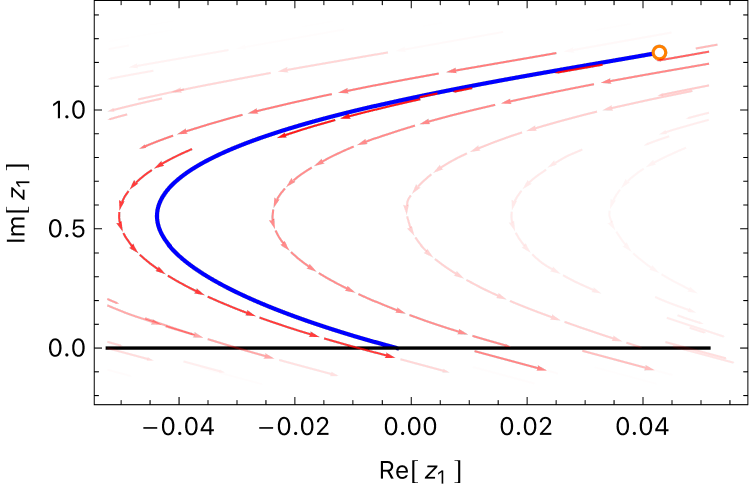}
    \includegraphics[width=0.235\textwidth]{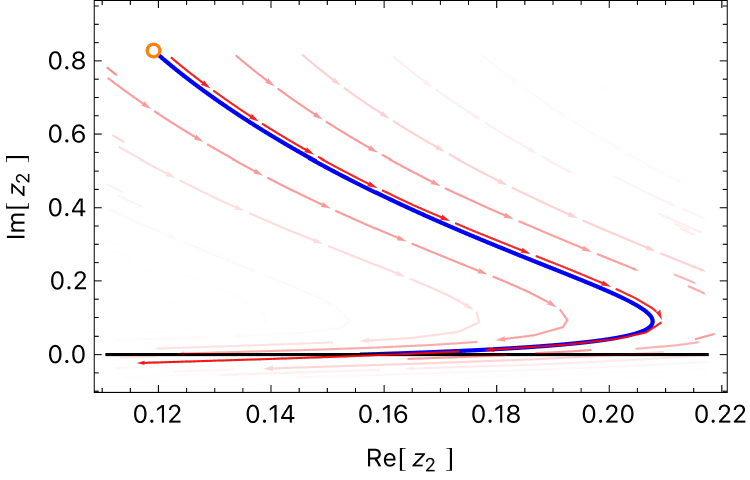}
    \caption{The upward flow from one of the saddle points to the original integration cycle for $\alpha=3.01$. Each panel shows the projection on the $(\Re z_i,\Im z_i)$-plane. The orange circle is the saddle point and the blue curve is the flow we obtained. We also show ${\rm d}z/{\rm d}s$ around the flow with red arrows.}
    \label{fig:airy_flow}
\end{figure}
\begin{figure}[t]
    \centering
    \includegraphics[width=0.4\textwidth]{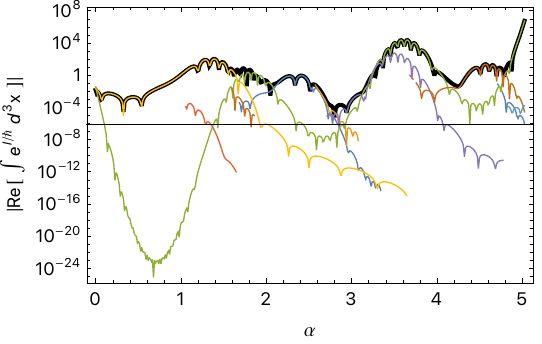}
    \caption{Comparison between the saddle-point approximation and direct numerical integration for the three-variable Airy-type integral. The colored lines represent contributions from individual saddle points, while the black thick line denotes results from direct numerical integration.}
    \label{fig:airy}
\end{figure}

For every $\alpha$, we solve the upward flow equations from each saddle point using the multiple shooting method. An example of the flow is shown in Fig.~\ref{fig:airy_flow}, where we take $\alpha=3.01$. It shows that the multiple shooting method successfully finds the upward flow even though the flow is quite complicated.
In Fig.~\ref{fig:airy}, we show the comparison between the numerical integration and the saddle-point approximation. The colored lines represent the contributions from each saddle point, while the black thick line denotes the results from direct numerical integration. We only show the absolute value of the real part, but we confirmed that the imaginary part is similar and the sign of the integral value agrees with the numerical one. We also do not show $\alpha\gtrsim5$ as the contribution from the green saddle becomes gigantic. The results from the saddle-point approximation exhibit excellent agreement with those from numerical integration, validating our method.
We also observe the Stokes phenomena where the colored lines end abruptly. This is because the thimble decomposition changes when crossing a Stokes line and the saddle point no longer contributes to the integral. More details are in Supplemental Material.

\subsection{Double-well potential}
Let us move on to a physically motivated case: the path integral in quantum mechanics with a double-well potential. The structure of its saddle points in the continuum limit is well understood~\cite{Tanizaki:2014xba}. However, determining which complex saddle points contribute to the path integral has remained a nontrivial open problem, except in trivial cases such as those corresponding to saddles lying on the original integration cycle $\mathcal Y$ or those with a positive real part in the exponent. Our method resolves this issue, by enabling a systematic identification of the relevant upward flows and the associated intersection numbers in the discretized version of this model, providing new insights into the contributions of nontrivial saddles in real-time path integrals.

\begin{figure}[t]
    \centering
    \includegraphics[width=0.235\textwidth]{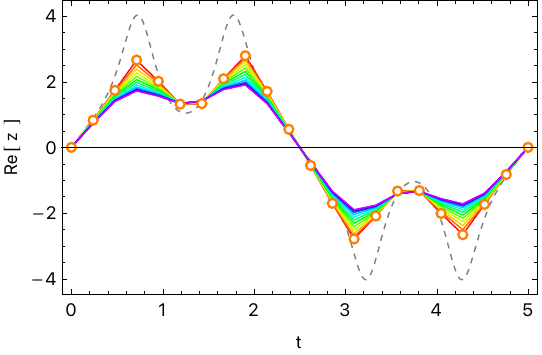}
    \includegraphics[width=0.235\textwidth]{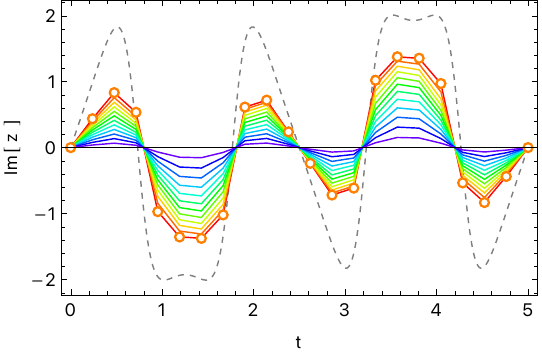}
    \caption{Upward flow from the saddle point $(n,m) = (4,2)$ to the original integration cycle. The black dashed line denotes the saddle point in the continuum limit, while the orange circles indicate the discretized saddle for $L = 20$. The upward flow trajectory is shown as colored lines, progressing from red ($s = 0$) to purple ($s = s_f$), with points plotted every $25$ steps.}
    \label{fig:dw_flow}
\end{figure}

As a model of the infinite-dimensional path integral, we discretize the time duration $T$ into $L$ segments with a lattice spacing $\Delta t=T/(L+1)$.
Then, the integration variables are the internal points, $x_i=x((i+1)\Delta t)$ with $i=0,\cdots,L-1$.
The exponent for a double-well potential is given by
\begin{align}
    \mathcal I(x)&=i\left[\frac12\sum_{i=1}^{L-1}\ab(\frac{x_i-x_{i-1}}{\Delta t})^2\Delta t+\frac{x_0^2+x_{L-1}^2}{2\Delta t}\right.\nonumber\\
    &\hspace{5ex}\left.-\frac12\sum_{i=0}^{L-1}(x_i^2-1)^2\Delta t-\Delta t\right]+\Delta\mathcal I(x).
\end{align}
Here, the boundary conditions are chosen as $x(0) = 0$ and $x(T) = 0$, corresponding to the computation of quantum corrections at the top of the potential hill.
The treatment of tunneling boundary conditions ($x(0) = -1$ and $x(T) = 1$) presents additional technical challenges beyond the scope of the present analysis and will be addressed in future work.
To ensure the saddle points are generic, we introduced a Morsification term to the action:
\begin{align}
    \Delta\mathcal I(x)&=ic\left[\sum_{i=0}^{L-1}\ab(1+\ab(\frac{i+1}{L+1})^2)x_i\right]\Delta t,\label{eq:morsification}
\end{align}
where $c$ is a small complex constant. Here, a non-trivial $i$-dependence is necessary to break all symmetries so that all saddle points become generic, which enables us to use Morse theory.
In this analysis, we set $T = 5$, $N=300$, $\delta r=0.01$ and $c=0.001+0.001i$.
For the saddle points analyzed below, we find that $c$ does not affect the intersection numbers provided $|c|$ is sufficiently small. This means that $c=0$ is not on the Stokes lines for these saddles.
In the continuum limit, all saddle points are given by the Jacobi elliptic functions and can be systematically labeled by two independent winding numbers $(n,m)$ counting how many times the solution winds around the torus \cite{Tanizaki:2014xba}.
We determine the corresponding saddle points in the discretized model numerically, and solve the upward flow equations from each saddle point using the multiple shooting method. Figure~\ref{fig:dw_flow} illustrates a representative example of the upward flow from the saddle point $(n,m)=(4,2)$ for $L=20$, demonstrating that the multiple shooting method reliably identifies the relevant upward flows even for large $L$. The maximum attainable value of $L$ is determined by the numerical stability and precision, and varies depending on the specific saddle point and the total time interval $T$.

We confirm that trivial cases, namely real saddles and saddles with positive real parts, consistently exhibit the correct intersection numbers. For non-trivial cases that have previously remained undetermined, our method enables the explicit determination of the intersection numbers, which we summarize for the first several saddle points in Table~\ref{table:dw_saddles}. Here, the sign of $n_\sigma$ depends on the orientation of the thimble, which we fix by $\Re(A_\sigma/(2\pi i\Delta t)^{\frac{L+1}{2}})>0$, where the denominator comes from the path integral measure. We confirmed that $n_\sigma$ is stable under variations of $L$.
Although our analysis relies on the discretized model, the stability under $L$ indicates that these results are likely to hold in the continuum limit.
These findings imply that multiple complex saddle points contribute to the real-time path integral, thereby validating previous theoretical analyses of complex saddle point contributions.

More details are in Supplemental Material.

\begin{table}[t]
    \centering
    \begin{tabular}{c|c|c|c|c|c|c}
        $n$&$m$&$\mathcal I_\infty[z]$&$\mathcal I(z)$&$L$&$n_\sigma$\\
        \hline
        $2$&$1$&$-1.280 + 1.427 i$&$-0.775 + 1.271 i$&$12$&$+1$\\
        $1$&$-2$&$-1.280 + 1.427 i$&$-0.764 + 1.257 i$&$12$&$+1$\\
        $3$&$2$&$-7.357 - 0.759 i$&$-$&$-$&$0$\\
        $2$&$-3$&$-7.357 - 0.759 i$&$-$&$-$&$0$\\
        $4$&$1$&$-14.926 + 19.727i$&$-5.783 + 17.860  i$&$16$&$-1$\\
        $1$&$-4$&$-14.926 + 19.727i$&$-5.783 + 17.862 i$&$16$&$-1$\\
        $4$&$2$&$-23.946 + 4.198 i$&$-15.311 + 6.545 i$&$20$&$-1$\\
        $2$&$-4$&$-23.946 + 4.198 i$&$-15.314 + 6.549  i$&$20$&$-1$\\
         $4$&$3$&$-21.025 - 18.980 i$&$-$&$-$&$0$\\
        $3$&$-4$&$-21.025 - 18.980 i$&$-$&$-$&$0$\\
    \end{tabular}
    \caption{The intersection numbers for the first few saddle points with $\Re I<0$ identified with $(n,m)$. Shown are also the values of $\mathcal I_\infty[z]$ obtained by integrating the continuous solution and $\mathcal I(z)$ the sum of the discretized and deformed action with Morsification parameter $c=0.001+0.001 i$.}
    \label{table:dw_saddles}
\end{table}
\section{Discussion and Conclusion}
In this paper, we have introduced a robust and efficient numerical method for determining intersection numbers of Lefschetz thimbles in multivariable settings. By applying the multiple shooting technique, we have overcome the challenges posed by the sensitivity to initial conditions inherent in the upward flow equations. Our method has been demonstrated to be effective in systems with up to tens of variables, achieving rapid convergence and high reliability. 
For reference, the computational efficiency of our approach is notable: On a single computational thread using a Rust implementation, the Airy-type example with $N = 200$ and $100$ Newton iterations typically completes in approximately $500\,\mathrm{ms}$. For the double-well example with $L = 20$, $N = 300$, and $200$ Newton iterations, the computation time is around $15\,\mathrm{s}$. These runtimes demonstrate the practicality of the method for high-dimensional problems.

We have applied our method to two representative systems: a three-variable Airy-type integral and a discretized path integral for a quantum mechanical system with a double-well potential. In both cases, we successfully identified the intersection numbers, including their signs. For the first case, the results from the saddle-point approximation showed excellent agreement with direct numerical integration, validating the accuracy of our approach. We also observed Stokes phenomena, where intersection numbers change, leading to abrupt shifts in the contributions from certain saddle points. In the second case, we determined the intersection numbers for several complex saddle points that had previously been undetermined, providing new insights into the contributions of complex saddles in real-time path integrals.

The methodology presented here is broadly applicable to a wide range of problems involving oscillatory integrals in physics and mathematics. The capability to efficiently compute intersection numbers in high-dimensional settings enables the investigation of complex systems that have previously been inaccessible to conventional approaches. While the current implementation is limited by double-precision arithmetic in the standard Newton's method, future enhancements, such as the adoption of quadruple precision or advanced variants of Newton's method, may further extend the feasible dimensionality.

Potential directions for future research include the extension of the method to cases involving degenerate saddle points and coinciding critical values. Additionally, exploring connections with complementary frameworks such as resurgence theory or exact WKB analysis could provide deeper insights into the structure of oscillatory integrals. Given the scalability of the approach, applications such as small lattice QCD models and cosmological models beyond minisuperspace represent promising avenues for further study.

\section*{Acknowledgments}
\begin{acknowledgments}
    This work is supported by the Slovenian Research Agency under the research core funding No. P1-0035 and in part by the research grant J1-4389.
\end{acknowledgments}

\bibliographystyle{apsrev4-2}
\bibliography{pl}

\clearpage
\onecolumngrid

\begin{center}
  \large \bf Supplemental Material
\end{center}

\section{Multiple shooting method}
In this section, we provide a detailed exposition of the multiple shooting method and its implementation in the context of upward flow equations. Specifically, we solve the flow equation with a normalized flow time,
\begin{equation}
    \pdv{z_i(s)}{s} = \frac{\overline{\pdv{\mathcal I}{z_i}}}{\left|\pdv{\mathcal I}{z}\right|},\label{eq:upward_flow}
\end{equation}
and appropriate boundary conditions. In the following, we use the notation $Z=(\Re z_0~\cdots~\Re z_{L-1}~\Im z_0~\cdots~\Im z_{L-1})^t$.

The multiple shooting method constructs a solution for $0 \leq s \leq (N-1)\delta s$ by partitioning the integration interval into $N-1$ subintervals:
\begin{equation}
    Z(s)=
    \begin{cases}
        \Phi(Z^{(0)};s)&(0\leq s<\delta s)\\
        \Phi(Z^{(1)};s-\delta s)&(\delta s\leq s<2\delta s)\\
        \vdots\\
        \Phi(Z^{(N-2)};s-(N-2)\delta s)&((N-2)\delta s\leq s<(N-1)\delta s)\\
        Z^{(N-1)}&s=(N-1)\delta s
    \end{cases}.
\end{equation}
Here, $\Phi(Z_{\rm init};s)$ denotes an exact solution of Eq.~\eqref{eq:upward_flow} with $\Phi(Z_{\rm init};0)=Z_{\rm init}$, {\it i.e.} it satisfies
\begin{equation}
    \pdv{\Phi}{s}(Z_{\rm init};s)=\frac{1}{\ab|\pdv{\mathcal I}{z}(\Phi(Z_{\rm init};s))|}
    \begin{pmatrix}
        \Re\pdv{\mathcal I}{z}(\Phi(Z_{\rm init};s))\\
        -\Im\pdv{\mathcal I}{z}(\Phi(Z_{\rm init};s))
    \end{pmatrix}.
\end{equation}
The number of subintervals $N-1$ is fixed in advance, while the step size $\delta s$ is treated as an optimization variable and determined self-consistently as part of the solution.
Due to the smallness of $\delta s$, the solution $\Phi$ remains stable and does not exhibit chaotic behavior within each subinterval.
Then, the continuity conditions are imposed as
\begin{equation}
    R^{(k)} = Z^{(k)} - \Phi(Z^{(k-1)}; \delta s) = 0,\label{eq:resc}
\end{equation}
for $k = 1, \ldots, N-1$. These conditions are solved together with the boundary conditions:
\begin{align}
    R^{(0)}_A&=|Z^{(0)}-Z_\sigma|-\delta r=0,\label{eq:res0p}\\
    R^{(0)}_B&=(W^-)^t(Z^{(0)}-Z_\sigma)=0,\label{eq:res0}\\
    R^{(N)}_B&=(0_{L\times L}~1_{L\times L})Z^{(N-1)}=0.\label{eq:resn}
\end{align}
Here, $W^\pm=(W_0^\pm~\cdots~W_{L-1}^\pm)$ denotes the eigenvectors of the Hessian matrix around the saddle point,
\begin{equation}
    \begin{pmatrix}
        \Re\odv[order=2]{\mathcal I}{z}&-\Im\odv[order=2]{\mathcal I}{z}\\
        -\Im\odv[order=2]{\mathcal I}{z}&-\Re\odv[order=2]{\mathcal I}{z}
    \end{pmatrix}W_i^\pm=\pm\lambda_iW_i^\pm,
\end{equation}
with $\lambda_i>0$.
The multiple shooting method reformulates the boundary value problem as a nonlinear system. Simultaneously enforcing the continuity conditions in Eq.~\eqref{eq:resc} and the boundary conditions in Eqs.~\eqref{eq:res0p}--\eqref{eq:resn}, the multiple shooting method yields a system of $2NL+1$ nonlinear equations,
\begin{equation}
    \mathcal R=
    \begin{pmatrix}
        R^{(0)}_A\\
        R^{(0)}_B\\
        R^{(1)}\\
        \vdots\\
        R^{(N-1)}\\
        R^{(N)}_B
    \end{pmatrix}=0,
\end{equation}
to be solved for the $2NL+1$ variables,
\begin{equation}
    \mathcal X=
    \begin{pmatrix}
        Z^{(0)}\\
        \vdots\\
        Z^{(N-1)}\\
        \delta s
    \end{pmatrix}.
\end{equation}

This nonlinear system can be efficiently solved using Newton's method, which typically exhibits rapid convergence in practice. In scenarios where Newton's method encounters difficulties, alternative approaches such as the steepest descent method may also be combined to ensure robust convergence.

The intersection number is then determined by whether Newton's method converges. As we will see, the convergence of Newton's method is very fast and we can clearly determine the intersection number for most of the cases. In this work, we restrict our attention to the generic case where $|n_\sigma| = 0$ or $1$. Situations with $|n_\sigma| > 1$ typically arise only in the presence of additional symmetries or degeneracies. In principle, all intersection points can be systematically identified by varying the initial guess in the Newton's method, although such cases are beyond the scope of the present analysis.

\subsection{Newton's method}
Given that the number of constraints matches the number of variables, Newton's method proceeds by solving the linear system
\begin{align}
    \pdv{\mathcal R}{\mathcal X}\Delta \mathcal X = -\mathcal R,\label{eq:newton}
\end{align}
and updating the variables according to
\begin{equation}
    \mathcal X \to \mathcal X + \Delta \mathcal X.
\end{equation}
Since the Jacobian $\partial \mathcal R/\partial \mathcal X$ is a $(2NL+1)\times(2NL+1)$ matrix, direct inversion is computationally intensive, scaling as $\mathcal{O}(N^3L^3)$. However, due to the nearly block-diagonal structure of the system, a solution can be efficiently obtained by exploiting this sparsity.

In this subsection, we present a streamlined implementation of Newton's method, which offers computational efficiency but may provide limited diagnostic information in cases where convergence is not achieved. For enhanced robustness and improved diagnostics, a more sophisticated approach is detailed in the following subsection.

We first address the continuity conditions, Eq.~\eqref{eq:resc}. The left-hand side of Eq.~\eqref{eq:newton} evaluates to
\begin{equation}
    \pdv{R^{(k)}}{\mathcal X}\Delta \mathcal X = \Delta Z^{(k)} - \pdv{\Phi}{Z}(Z^{(k-1)}; \delta s)\Delta Z^{(k-1)} - \pdv{\Phi}{s}(Z^{(k-1)}; \delta s)\Delta \delta s,
\end{equation}
which yields the recurrence relation
\begin{equation}
    \Delta Z^{(k)} = J_Z^{(k-1)}\Delta Z^{(k-1)} + J_s^{(k-1)}\Delta \delta s - R^{(k)}, \label{eq:recursive}
\end{equation}
where
\begin{align}
    J_Z^{(k)} = \pdv{\Phi}{Z}(Z^{(k)}; s), \qquad J_s^{(k)} = \pdv{\Phi}{s}(Z^{(k)}; \delta s).
\end{align}
Here, the derivatives are computed numerically as
\begin{align}
    \pdv{\Phi}{Z_i}(Z; s) &\simeq \frac{\Phi(Z + h v_i; s) - \Phi(Z - h v_i; s)}{2h}, \\
    \pdv{\Phi}{s}(Z; s) &\simeq \frac{\Phi(Z; s + h) - \Phi(Z; s - h)}{2h},
\end{align}
with $(v_i)_j = \delta_{ij}$ and a small $h > 0$.

By iterating Eq.~\eqref{eq:recursive}, each $\Delta Z^{(k)}$ can be expressed in terms of $\Delta Z^{(0)}$ and $\Delta \delta s$:
\begin{align}
    \Delta Z^{(k)} = \mathbf{J}_Z^{(k)} \Delta Z^{(0)} + \mathbf{J}_s^{(k)} \Delta \delta s - \mathbf{R}^{(k)}, \label{eq:bigj}
\end{align}
where the coefficients satisfy the recursion relations
\begin{align}
    \mathbf{J}_Z^{(k)} &= J_Z^{(k-1)} \mathbf{J}_Z^{(k-1)}, \\
    \mathbf{J}_s^{(k)} &= J_Z^{(k-1)} \mathbf{J}_s^{(k-1)} + J_s^{(k-1)}, \\
    \mathbf{R}^{(k)}   &= J_Z^{(k-1)} \mathbf{R}^{(k-1)} + R^{(k)},
\end{align}
with initial conditions
\begin{align}
    \mathbf{J}_Z^{(1)} = J_Z^{(0)}, \qquad
    \mathbf{J}_s^{(1)} = J_s^{(0)}, \qquad
    \mathbf{R}^{(1)} = R^{(1)}.
\end{align}

We now proceed to the boundary conditions. We define $\Delta Z^{(0)}_\pm$ via
\begin{equation}
    \Delta Z^{(0)}=W^+\Delta Z^{(0)}_++W^-\Delta Z^{(0)}_-.
\end{equation}
Then, the left-hand side of Eq.~\eqref{eq:newton} for $R_B^{(0)}$ evaluates to
\begin{equation}
    \pdv{R_B^{(0)}}{\mathcal X}\Delta\mathcal X = \Delta Z^{(0)}_-,
\end{equation}
which immediately yields the relation
\begin{equation}
    \Delta Z^{(0)}_- = -R^{(0)}_B.
\end{equation}

Next, the left-hand side of Eq.~\eqref{eq:newton} for $R_B^{(N)}$ and $R^{(0)}_A$ are given by
\begin{align}
    \pdv{R^{(N)}_B}{\mathcal X}\Delta\mathcal X&=(0_{L\times L}~1_{L\times L})\Delta Z^{(N-1)}\nonumber\\
    &=(0_{L\times L}~1_{L\times L})\ab(\mathbf J_Z^{(N-1)}\Delta Z^{(0)}+\mathbf J_s^{(N-1)}\Delta \delta s-\mathbf R^{(N-1)})\nonumber\\
    &=(0_{L\times L}~1_{L\times L})\ab(\mathbf J_Z^{(N-1)}W^+\Delta Z^{(0)}_++\mathbf J_s^{(N-1)}\Delta \delta s-\mathbf R^{(N-1)}-\mathbf J_Z^{(N-1)}W^-R_B^{(0)}),
\end{align}
and
\begin{equation}
    \pdv{R^{(0)}_A}{\mathcal X}\Delta\mathcal X=\frac{Z^{(0)}-Z_\sigma}{|Z^{(0)}-Z_\sigma|}\Delta Z^{(0)}.
\end{equation}
Hereafter, $1_{n\times n}$ indicates an $n\times n$ unit matrix and $0_{n\times m}$ indicates an $n\times m$ zero matrix. We also use $0_n$ for a zero vector.
These yield the linear system
\begin{align}
    \begin{pmatrix}
        (0_{L\times L}~1_{L\times L})\mathbf J_Z^{(N-1)}W^+&(0_{L\times L}~1_{L\times L})\mathbf J_s^{(N-1)}\\
        \frac{Z^{(0)}-Z_\sigma}{|Z^{(0)}-Z_\sigma|}W^+&0
    \end{pmatrix}
    \begin{pmatrix}
        \Delta Z^{(0)}_+\\
        \Delta \delta s
    \end{pmatrix}
    =
    \begin{pmatrix}
        (0_{L\times L}~1_{L\times L})\ab(\mathbf R^{(N-1)}+\mathbf J_Z^{(N-1)}W^-R_B^{(0)})-R_B^{(N)}\\
        \frac{Z^{(0)}-Z_\sigma}{|Z^{(0)}-Z_\sigma|}W^-R_B^{(0)}-R_A^{(0)}
    \end{pmatrix}.
\end{align}
Since the matrix in the left-hand side is an $(L+1)\times(L+1)$ matrix, we can solve the linear system and obtain $\Delta Z^{(0)}$ and $\Delta \delta s$. Then, all $\Delta Z^{(k)}$'s are calculated from Eq.~\eqref{eq:recursive} or \eqref{eq:bigj}.

After obtaining $\Delta \mathcal X$, we update $\mathcal X$. However, a direct update may not always lead to a reduction in the residual $|\mathcal R(\mathcal X)|$ even around a solution. This happens when the Jacobian becomes ill-conditioned or rapidly changes around the solution. To ensure convergence, we implement a line search strategy. Specifically, we introduce a parameter $0 < \alpha \leq 1$ and $0 < c_{\rm LS}$ such that
\begin{equation}
    \mathcal R(\mathcal X + \alpha \Delta \mathcal X)<c_{\rm LS}\mathcal R(\mathcal X).
\end{equation}
In practice, we initially set $c_{\rm LS} > 1$ to allow the algorithm to explore the solution space and approach a region near the solution. Once the residual is sufficiently reduced, we switch to $c_{\rm LS} \leq 1$ to ensure stable convergence and prevent divergence of the update step. If a suitable $\alpha$ is found, the variables are updated as
\begin{equation}
    Z^{(k)} \to Z^{(k)} + \alpha \Delta Z^{(k)}, \quad \delta s \to |\delta s + \alpha \Delta \delta s|.
\end{equation}
Here, we take the absolute value of $\delta s$ to ensure it remains positive.

If no suitable $\alpha$ is found for $\alpha>0.001$, this generally indicates either the absence of a solution in the vicinity or that the Jacobian is ill-conditioned. In such situations, the variables are updated using a small value of $\alpha$, enabling the algorithm to explore another region.

\subsection{Newton's method with diagnosis}
There are several reasons that Newton's method fails. One is simply that there is no solution. Another is that multiplication of many $J_Z^{(k)}$ makes the vectors too squeezed so that we cannot fully recover the information of tangent vectors.
A relatively rare case is that there is a flat direction, which appears when the flow does not transverse but only touches the $\Im z=0$ plane. Lastly, we observed that the Jacobian matrix typically becomes ill-conditioned when the saddle point is close to another saddle point.
To systematically diagnose potential issues, we analyze the propagation of the coordinates $(\Delta Z^{(0)}, \Delta \delta s)$ to those of $\Delta Z^{(N-1)}$. This allows us to monitor the evolution of tangent vectors and identify cases where information is lost due to ill-conditioning or degeneracies in the Jacobian. We present the details below and we use this version of Newton's method for all analyses.

Since $(\Delta Z^{(0)}, \Delta \delta s)$ over-completes the $2L$-dimensional surface, we eliminate one degree of freedom by selecting an index $i = a$ that maximizes $|(Z^{(0)} - Z_\sigma) \cdot W^+_a|$. Then, Eq.~\eqref{eq:newton} for the anchor condition $R^{(0)}_A$ is solved for $\Delta Z^{(0)}_{+a}$ as follows:
\begin{equation}
    \Delta Z^{(0)}_{+a} = \mathcal N_a \left[ \frac{(Z^{(0)} - Z_\sigma)^t}{|Z^{(0)} - Z_\sigma|} W^- R_B^{(0)} - R_A^{(0)} - \frac{(Z^{(0)} - Z_\sigma)^t}{|Z^{(0)} - Z_\sigma|} \tilde W^+ \Delta \tilde Z_+^{(0)} \right],
\end{equation}
where $\tilde W^+$ denotes the matrix $W^+$ with the $a$-th column removed, and $\Delta \tilde Z_+^{(0)}$ is the vector $\Delta Z_+^{(0)}$ with the $a$-th component omitted. Here, $\mathcal N_a$ is given by
\begin{equation}
    \mathcal N_a = \frac{|Z^{(0)} - Z_\sigma|}{(Z^{(0)} - Z_\sigma)^t W^+_a}.
\end{equation}

We now express $\Delta Z^{(1)}$ in terms of the reduced coordinate system. Let $i = b$ the index maximizing $|\rho \cdot W^-_b|$, where
\begin{equation}
    \rho = -J_Z^{(0)} W^- R_B^{(0)} + \mathcal N_a J_Z^{(0)} W^+_a \left[ \frac{(Z^{(0)} - Z_\sigma)^t}{|Z^{(0)} - Z_\sigma|} W^- R_B^{(0)} - R_A^{(0)} \right] - R^{(1)}.
\end{equation}
Then, $\Delta Z^{(1)}$ can be written as
\begin{align}
    \Delta Z^{(1)} = J_Z^{(0)} \tilde W^+ \Delta \tilde Z_+^{(0)}
    - \mathcal N_a J_Z^{(0)} W^+_a \frac{(Z^{(0)} - Z_\sigma)^t}{|Z^{(0)} - Z_\sigma|} \tilde W^+ \Delta \tilde Z_+^{(0)}
    + J_s^{(0)} \Delta \delta s + \rho.
\end{align}

To facilitate further analysis, we introduce the notation
\begin{equation}
    \Delta Z^{(1)} = \mathbf{J}^{(1)} \Delta V,
\end{equation}
with
\begin{align}
    \mathbf{J}^{(1)} &=
    \begin{pmatrix}
        J_Z^{(0)} \tilde W^+ - \mathcal N_a J_Z^{(0)} W^+_a \frac{(Z^{(0)} - Z_\sigma)^t}{|Z^{(0)} - Z_\sigma|} \tilde W^+ & J_s^{(0)} & \rho & J_Z^{(0)} \tilde W^-
    \end{pmatrix}, \\
    \Delta V &=
    \begin{pmatrix}
        \Delta \tilde Z_+^{(0)} \\
        \Delta \delta s \\
        1 \\
        0_{L-1}
    \end{pmatrix}.
\end{align}
Here, $\tilde W^-$ denotes $W^-$ with the $b$-th column removed.

For the analysis of the tangent space propagation, we perform a QR decomposition of $\mathbf J^{(1)}$:
\begin{equation}
    \mathbf J^{(1)} = \mathtt{Q}^{(1)} \mathtt{R}^{(1)},
\end{equation}
where $\mathtt{Q}^{(1)}$ is an orthogonal matrix and $\mathtt{R}^{(1)}$ is an upper triangular matrix.
For $k \geq 2$, we iteratively express $\Delta Z^{(k)}$ in terms of $\Delta V$:
\begin{align}
    \Delta Z^{(k)}=\mathbf J^{(k)}\Delta V,
\end{align}
where
\begin{align}
    \mathbf J^{(k)}&=J_Z^{(k-1)}\mathtt{Q^{(k-1)}}\mathtt{R^{(k-1)}}+
    \begin{pmatrix}
        0_{2L\times L-1}&J_s^{(k-1)}&-R^{(k)}&0_{2L\times L-1}
    \end{pmatrix}\nonumber\\
    &=\ab[J_Z^{(k-1)}\mathtt{Q^{(k-1)}}+
    \begin{pmatrix}
        0_{2L\times L-1}&J_s^{(k-1)}&-R^{(k)}&0_{2L\times L-1}
    \end{pmatrix}(\mathtt{R^{(k-1)}})^{-1}]\mathtt{R^{(k-1)}}\nonumber\\
    &=\ab[\mathtt{Q^{(k)}}\mathtt{\tilde R^{(k)}}]\mathtt{R^{(k-1)}}\nonumber\\
    &=\mathtt{Q^{(k)}}\mathtt{R^{(k)}}.
\end{align}
Since the multiplication of triangle matrices are a triangle matrix, $\mathtt{R^{(k)}}\Delta V$ has non-zero elements only in upper $L+1$ components. This decomposition enables a stable and systematic representation of the basis vectors, with $\mathtt{Q}^{(k)}$ providing an orthonormal basis and $\mathtt{R}^{(k)}$ encoding the coordinate transformation.

Finally, we solve
\begin{equation}
    (0_{L\times L}~1_{L\times L})\mathtt{Q^{(N-1)}}\mathtt{R^{(N-1)}}\Delta V=-R_B^{(N)},
\end{equation}
or more explicitly,
\begin{equation}
    \ab[(0_{L\times L}~1_{L\times L})\mathtt{Q^{(N-1)}}\mathtt{R^{(N-1)}}
    \begin{pmatrix}
    1_{L\times L}\\
    0_{L\times L}
    \end{pmatrix}]
    \begin{pmatrix}
        \Delta \tilde Z^{(0)}_+\\
        \Delta \delta s
    \end{pmatrix}=
    -R_B^{(N)}-(0_{L\times L}~1_{L\times L})\mathtt{Q^{(N-1)}}\mathtt{R^{(N-1)}}
    \begin{pmatrix}
        0_{L}\\
        1\\
        0_{L-1}
    \end{pmatrix}.
\end{equation}

We now introduce diagnostic matrices to assess the quality and transversality of the propagated tangent space. Define
\begin{align}
    \mathtt{Q_F} &=
    \begin{pmatrix}
        0_{L\times L} & 1_{L\times L}
    \end{pmatrix}
    \mathtt{Q}^{(N-1)}
    \begin{pmatrix}
        1_{L\times L} \\
        0_{L\times L}
    \end{pmatrix}, \\
    \mathtt{R_F} \mathtt{D} &=
    \begin{pmatrix}
        1_{L\times L} & 0_{L\times L}
    \end{pmatrix}
    \mathtt{R}^{(N-1)}
    \begin{pmatrix}
        1_{L\times L} \\
        0_{L\times L}
    \end{pmatrix},
\end{align}
where $\mathtt{D}$ is a diagonal normalization matrix chosen such that the diagonal elements of $\mathtt{R_F}$ are unity.

The matrix $\mathtt{Q_F}$ characterizes the transversality of the flow with respect to the $\Im z = 0$ plane. If $\mathtt{Q_F}$ possesses a zero eigenvalue, the flow merely touches, rather than transverses, the plane, indicating a degenerate intersection. The conditioning of $\mathtt{R_F}$ reflects the numerical stability of the propagated tangent vectors: if $||\mathtt{R_F}||_{\max}$ approaches the limits of double precision (e.g., $10^{15}$), the tangent vectors are excessively squeezed, and the resolution of the intersection becomes unreliable.

For further verification, one may explicitly compute the inverses $\mathtt{Q_F}^{-1}\mathtt{Q_F}$ and $\mathtt{R_F}^{-1}\mathtt{R_F}$ to check their proximity to the identity matrix. Significant deviations from the identity, in the absence of the aforementioned pathologies, indicate ill-conditioning arising from other sources. This happens, for example, when two saddle points are close to each other and the structure of the flow becomes complicated between these saddle points. In such a case, increasing $\delta r$ or $N$ helps the convergence of Newton's method.

\subsection{Improved anchor condition}
When two saddle points are in close proximity, the flow structure in their vicinity becomes highly intricate, often resulting in degraded convergence of Newton's method. To address this challenge, we introduce an improved anchor condition that enhances numerical stability in such regimes.

The original anchor condition, Eq.~\eqref{eq:res0p}, constrains the distance from the saddle point $Z_\sigma$ using the Euclidean norm. However, this approach does not account for the anisotropic expansion rates along different directions in the tangent space. To better capture the geometry of the flow, we incorporate the eigenvector and eigenvalue information encoded in $W$ and $\Lambda$.

Specifically, we define a modified anchor condition as
\begin{equation}
    \tilde R_A^{(0)} = \left| W \Lambda^q W^t (Z^{(0)} - Z_\sigma) \right| - \delta r = 0,
\end{equation}
where
\begin{equation}
    W = \begin{pmatrix} W^- & W^+ \end{pmatrix},
\end{equation}
and
\begin{equation}
    \Lambda = \mathrm{diag}\left( \frac{\lambda_0}{\lambda_{\min}}, \ldots, \frac{\lambda_{L-1}}{\lambda_{\min}}, \frac{\lambda_0}{\lambda_{\min}}, \ldots, \frac{\lambda_{L-1}}{\lambda_{\min}} \right).
\end{equation}
Here, $\lambda_{\min}$ denotes the smallest eigenvalue among $\lambda_i$, and $q > 0$ is a tunable parameter that controls the weighting of directions according to their expansion rates. In practice, we find that $1\lesssim q\lesssim2$ provides robust performance for the Airy-type example and there was no preferred $q$ for the double well example. We use $q=1.5$ for all the analyses in this paper.
\section{Sign of intersection number}
The intersection number is defined via the intersection pairing $n_\sigma = \langle \mathcal Y, \mathcal K_\sigma \rangle$, with the sign convention fixed by $\langle \mathcal J_\sigma, \mathcal K_\tau \rangle = \delta_{\sigma\tau}$. The multiple shooting method not only identifies the intersection points but also systematically propagates the tangent space from the saddle point to the intersection, thereby enabling a precise and consistent determination of the sign of each intersection number.

Upon successful convergence of the multiple shooting method, the relation $Z^{(k+1)} = \Phi(Z^{(k)}; \delta s)$ holds for each subinterval, leading to
\begin{equation}
    Z^{(N-1)} = \Phi(\Phi(\cdots \Phi(Z^{(0)}; \delta s) \cdots; \delta s); \delta s).
\end{equation}
This enables the propagation of tangent vectors from $\delta Z^{(0)} \in T_{Z^{(0)}} \mathcal K_\sigma$ to $\delta Z^{(N-1)} \in T_{Z^{(N-1)}} \mathcal K_\sigma$ via infinitesimal variations:
\begin{align}
    \delta Z^{(N-1)} &= \pdv{\Phi}{Z}(Z^{(N-2)}; \delta s) \cdots \pdv{\Phi}{Z}(Z^{(0)}; \delta s) \delta Z^{(0)} \nonumber \\
    &= \mathbf{J}_Z^{(N-1)} \delta Z^{(0)}.
\end{align}

Since $Z^{(0)}$ is chosen in close proximity to the saddle point $Z_\sigma$, we approximate $T_{Z_\sigma} \mathcal K_\sigma \simeq T_{Z^{(0)}} \mathcal K_\sigma$. The tangent space $T_{Z^{(0)}} \mathcal K_\sigma$ is spanned by the columns of $W^+$, while the tangent space $T_{Z^{(0)}} \mathcal J_\sigma$ is spanned by the columns of $W^-$.
In the following, we fix the orientation of $\mathcal J_\sigma$ by the ordering of $W^-$: If we take the opposite orientation, the sign of the intersection number flips.
For each saddle point, we introduce an orientation factor $\Sigma = \pm 1$ arising from the ordering of $W^-$ and $W^+$, such that
\begin{equation}
    1 = \langle \mathcal J_\sigma, \mathcal K_\sigma \rangle = \Sigma \, \mathrm{sign} \det \begin{pmatrix}
        W^- & W^+
    \end{pmatrix}.
\end{equation}

At the intersection point $Z^{(N-1)}$, the tangent space $T_{Z^{(N-1)}} \mathcal Y$ is spanned by the first $L$ elements with the standard ordering. Thus, the sign of the intersection number is determined by
\begin{equation}
    \langle \mathcal Y, \mathcal K_\sigma \rangle = \Sigma \, \mathrm{sign} \det
        \begin{pmatrix}
            \begin{pmatrix}
                1_{L \times L} \\
                0_{L \times L}
            \end{pmatrix} & \mathbf{J}_Z^{(N-1)} W^+
        \end{pmatrix}.
\end{equation}

We now generalize the sign determination to the diagnostic version.
Recall that the flow map $\Phi(Z^{(k)}; s)$ satisfies the semigroup property,
\begin{equation}
    \Phi(Z^{(k)}; \delta s' + \delta s) = \Phi(\Phi(Z^{(k)}; \delta s'); \delta s),
\end{equation}
which implies the following relation for its derivatives:
\begin{align}
    \pdv{\Phi}{s}(Z^{(k)}; \delta s) &= \pdv{\Phi}{Z}(Z^{(k)}; \delta s) \pdv{\Phi}{s}(Z^{(k)}; 0) \nonumber \\
    &= \pdv{\Phi}{Z}(Z^{(k)}; \delta s) \pdv{\Phi}{s}(Z^{(k-1)}; \delta s).
\end{align}
Since $\partial \Phi/\partial s(Z^{(k-1)};\delta s)$ obeys the same propagation equation as $\delta Z^{(k)}$ starting from $\delta Z^{(1)}\in T_{Z^{(1)}}K_\sigma$, one of the tangent vectors of $T_{Z^{(1)}} \mathcal K_\sigma$ can be replaced by $J_s^{(0)}$. Approximating $T_{Z_\sigma} \mathcal K_\sigma \simeq T_{Z^{(1)}} \mathcal K_\sigma$, the tangent space at the saddle point is represented by the first $L$ columns of $\mathbf{J}^{(1)}$.
Accordingly, the intersection pairing at the saddle point is given by
\begin{equation}
    1 = \langle \mathcal J_\sigma, \mathcal K_\sigma \rangle = \Sigma' \, \mathrm{sign} \det
        \begin{pmatrix}
            W^- & \mathbf{J}^{(1)} \begin{pmatrix}
                1_{L \times L} \\
                0_{L \times L}
            \end{pmatrix}
        \end{pmatrix},
\end{equation}
where $\Sigma'$ is the orientation factor.

These tangent vectors are then propagated to the intersection point as
\begin{equation}
    \langle \mathcal Y, \mathcal K_\sigma \rangle = \Sigma' \, \mathrm{sign} \det
        \begin{pmatrix}
            \begin{pmatrix}
                1_{L \times L} \\
                0_{L \times L}
            \end{pmatrix} & \mathbf{J}^{(N-1)} \begin{pmatrix}
                1_{L \times L} \\
                0_{L \times L}
            \end{pmatrix}
        \end{pmatrix}.
\end{equation}
\section{Detailed analysis on Airy-type integral}
We provide a comprehensive analysis of the Airy-type example discussed in the main text. Throughout this study, we fix $\hbar = 0.05$, corresponding to an expected error of approximately $5\%$ in the saddle-point approximation. While smaller values of $\hbar$ would improve the accuracy of the approximation, they also increase the oscillatory behavior of the integrand, making direct numerical integration more challenging. For the multiple shooting method, we employ $N = 200$ subintervals, which is more than sufficient for most parameter regimes. However, in regions where saddle points are closely spaced, the local nonlinearity becomes significant, and a larger $N$ is required to achieve robust convergence of the algorithm. For $\Phi$, we approximate the solution with a single step of the Dormand-Prince method, which also provides an error estimate. We have verified that the error is small enough and introducing many steps does not significantly affect the results.

We set the separation parameter $\delta r = 0.01$ between $Z^{(0)}$ and the saddle point $Z_\sigma$. In instances where convergence is not achieved for smaller values of $\delta r$, we incrementally increase $\delta r$ up to $0.03$. This adjustment is particularly important when saddle points are in close proximity, as the resulting flow structure becomes highly intricate and the Jacobian may become ill-conditioned. Employing a larger $\delta r$ in such cases improves numerical stability and facilitates successful convergence.

We employ the following strategy for generating initial guesses for $Z^{(k)}$ and $\delta s$:
\begin{align}
    \log_{10}(\delta s) &\sim \mathcal{U}(-2, 0), \\
    \delta Z_i &\sim \mathcal{N}(0, 1), \\
    Z^{(0)} - Z_\sigma &= \frac{\delta Z}{|\delta Z|} \, \delta r, \\
    Z^{(k+1)} - Z^{(k)} &= \frac{\delta Z}{|\delta Z|} \, \delta s,
\end{align}
where $\mathcal{U}(a, b)$ denotes the uniform distribution over the interval $[a, b]$, and $\mathcal{N}(\mu, \sigma)$ denotes the normal distribution with mean $\mu$ and standard deviation $\sigma$. Here, we use the same $\delta Z$ across all subintervals $k$; we have also verified that allowing $\delta Z$ to vary with $k$ yields comparable convergence.

\begin{figure}[t]
    \centering
    \includegraphics[width=0.3\textwidth]{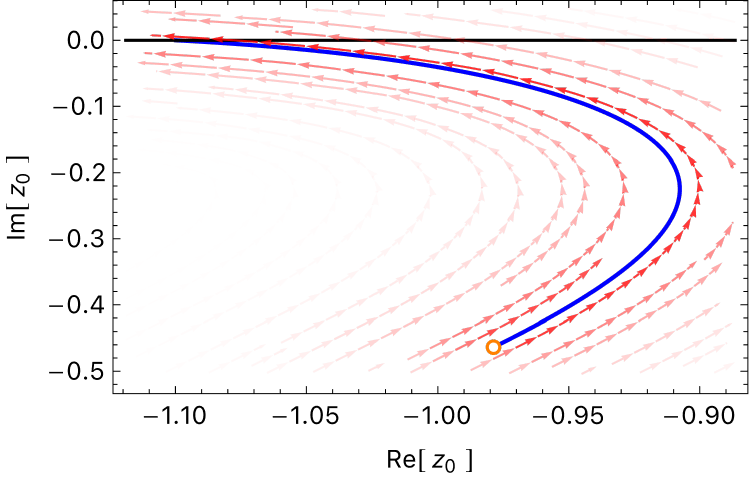}
    \includegraphics[width=0.3\textwidth]{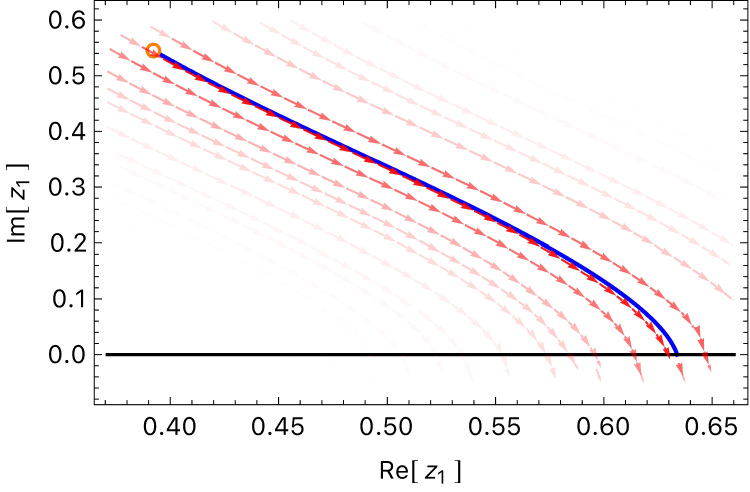}
    \includegraphics[width=0.3\textwidth]{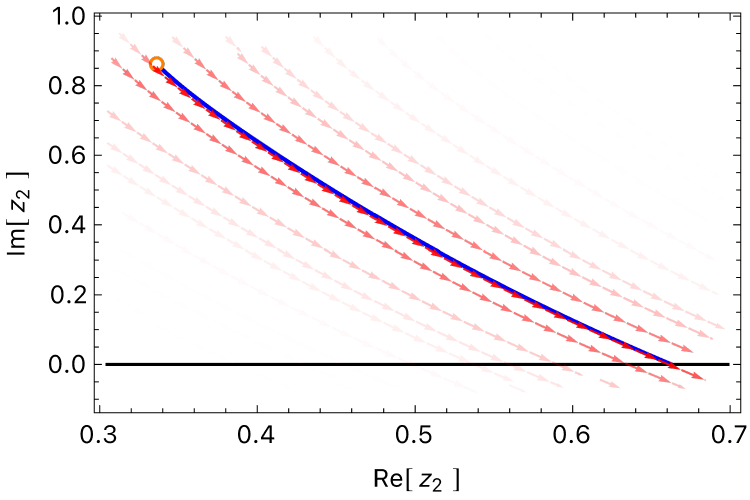}
    \includegraphics[width=0.3\textwidth]{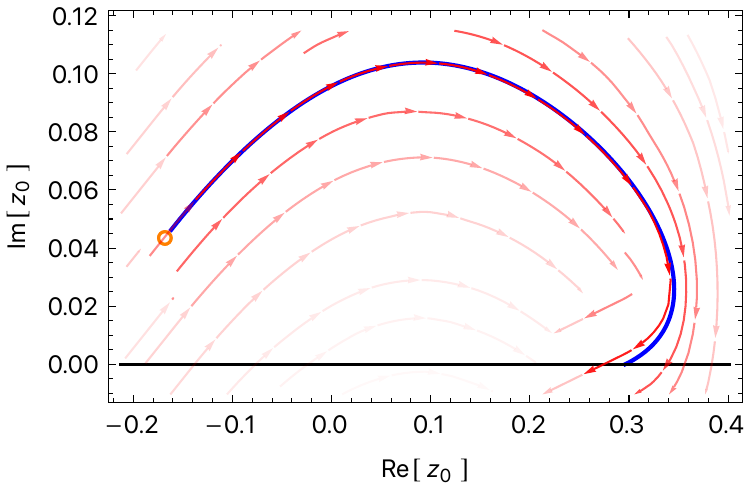}
    \includegraphics[width=0.3\textwidth]{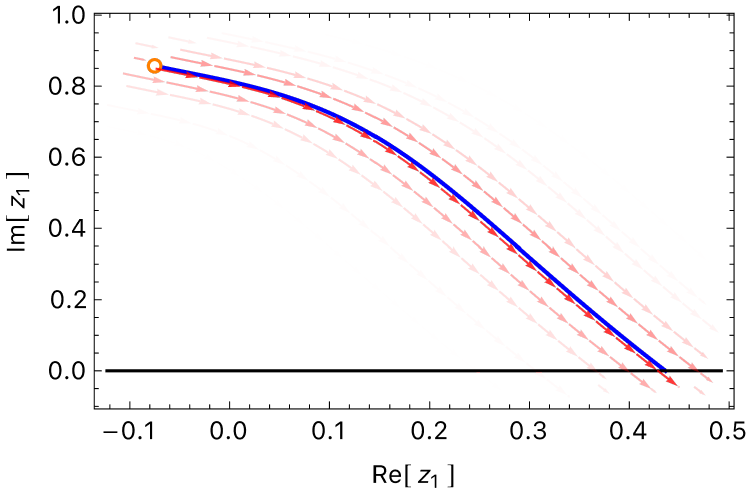}
    \includegraphics[width=0.3\textwidth]{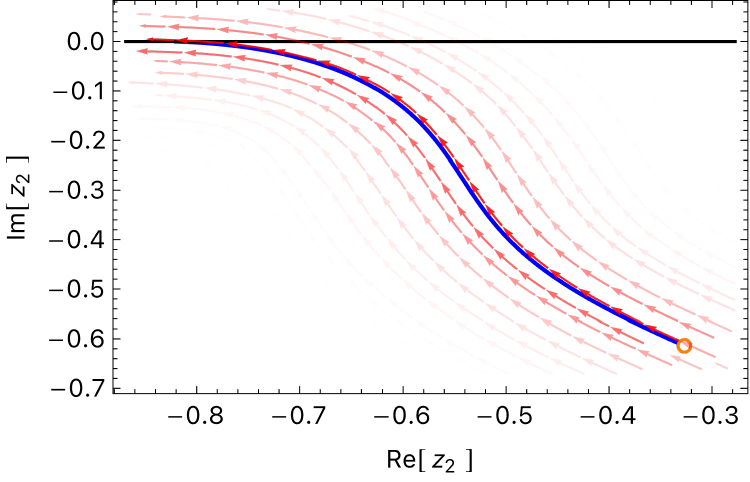}
    \caption{The same figure as Fig.~\ref{fig:airy_flow}, but for those used in the analysis of convergence. The top panels show a typical convergent case ($\alpha=1.6$, $\delta r=0.01$). The bottom panels show a special case where two saddle points are close and the Jacobian becomes ill-conditioned ($\alpha=2.6$, $\delta r=0.01$). The left, middle, and right panels show the $z_0$, $z_1$, and $z_2$ components of the flow, respectively.}
    \label{fig:airy_examples}
\end{figure}

\begin{figure}[t]
    \centering
    \includegraphics[width=0.3\textwidth]{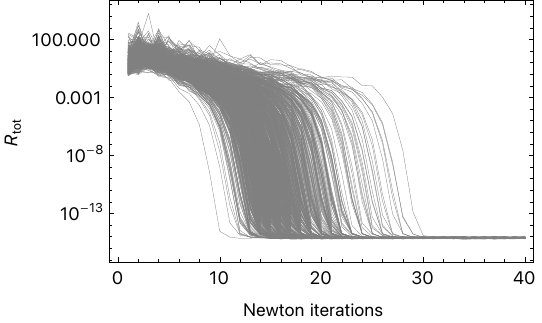}
    \includegraphics[width=0.3\textwidth]{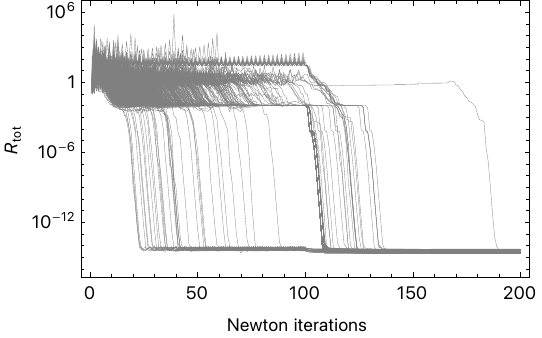}
    \includegraphics[width=0.3\textwidth]{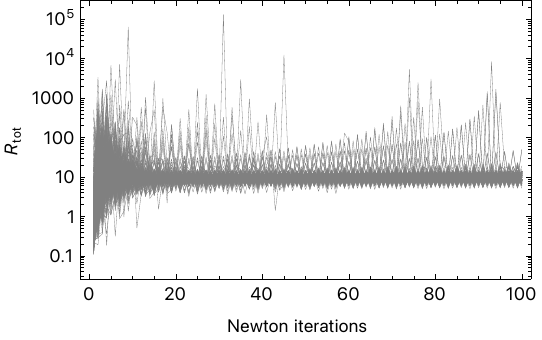}
    \caption{Convergence of Newton's method. The left panel shows a typical convergent case ($\alpha=1.6$, $\delta r=0.01$, the saddle shown in the top panels of Fig.~\ref{fig:airy_examples}). The middle panel shows a special case where two saddle points are close and the Jacobian becomes ill-conditioned ($\alpha=2.6$, $\delta r=0.01$, the saddle shown in  the bottom panels of Fig.~\ref{fig:airy_examples}). The right panel shows a non-convergent case where there is no solution ($\alpha=1.6$, $\delta r=0.01$, another saddle specified in the text).}
    \label{fig:airy_convergence}
\end{figure}

We now discuss the convergence properties of Newton's method in detail. To monitor the residual, we define
\begin{equation}
    \mathcal R_{\rm tot}=||\mathcal R||_{L^2}.
\end{equation}
For a typical scenario, we set $\alpha=1.6$ and $\delta r=0.01$, and do not employ line search. The saddle point and the flow is shown in the top panels of Fig.~\ref{fig:airy_examples}. The convergence behavior over $1000$ trials with different initial guesses is shown in the left panel of Fig.~\ref{fig:airy_convergence}. Each iteration requires approximately $5\,\mathrm{ms}$ on a single computational thread with a Rust implementation, and the method typically converges within $100\,\mathrm{ms}$. The residual $\mathcal R_{\rm tot}$ rapidly reaches $\sim 10^{-15}$, which is limited by double-precision arithmetic. 

In cases where two saddle points are in close proximity, the Jacobian may become ill-conditioned. For example, with $\alpha=2.6$ and $\delta r=0.01$, convergence requires the use of line search. The saddle point and the flow is shown in the bottom panels of Fig.~\ref{fig:airy_examples}. We first perform $100$ iterations without line search, followed by $100$ iterations with line search using $c_{\rm LS}=1$. The resulting convergence behavior is shown in the middle panel of Fig.~\ref{fig:airy_convergence}. The optimization stalls at $\mathcal R_{\rm tot} \simeq 0.01$, with $||\mathtt{R_F}||_{\max} \simeq 5.7 \times 10^{11}$, indicating that the propagated tangent vectors are excessively squeezed and the Jacobian matrix cannot be reliably inverted. Upon activating line search, Newton's method resumes convergence.

Finally, we consider a case where no solution exists, taking $\alpha=1.6$ and $\delta r=0.01$. We compute the saddle point around $z_0\simeq0.34-0.8i$, $z_1\simeq0.37-0.47i$, and $z_2\simeq-0.92+0.42i$. The right panel of Fig.~\ref{fig:airy_convergence} shows that $\mathcal R_{\rm tot}$ remains oscillatory and does not decrease below $\mathcal O(1)$, clearly distinguishing non-convergent behavior from the previous cases. We have verified that the Jacobian is well-conditioned in this regime by checking that $\mathtt{R_F}^{-1}\mathtt{R_F}$ and $\mathtt{Q_F}^{-1}\mathtt{Q_F}$ are close to the identity.

\begin{figure}[t]
    \centering
    \includegraphics[width=0.3\textwidth]{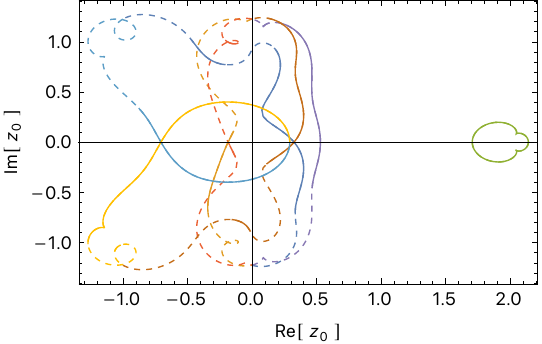}\hspace{1ex}
    \includegraphics[width=0.3\textwidth]{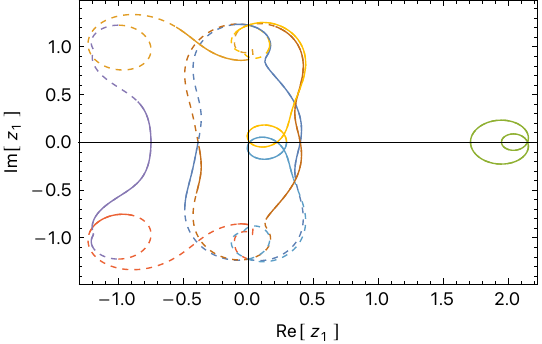}\hspace{1ex}
    \includegraphics[width=0.3\textwidth]{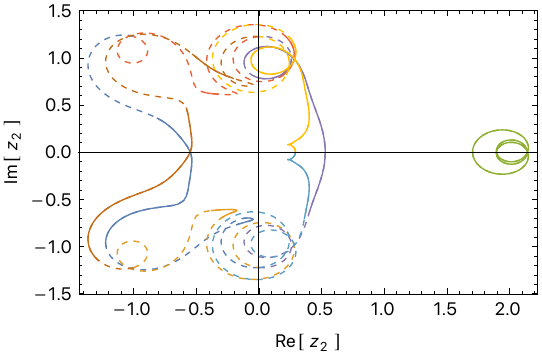}
    \caption{Trajectories of saddles as $\alpha$ goes from $0$ to $2\pi$. The solid lines indicate $|n_\sigma|=1$ and the dashed lines indicate $|n_\sigma|=0$.}
    \label{fig:airy_saddles}
\end{figure}

\begin{figure}[t]
    \centering
    \includegraphics[width=0.4\textwidth]{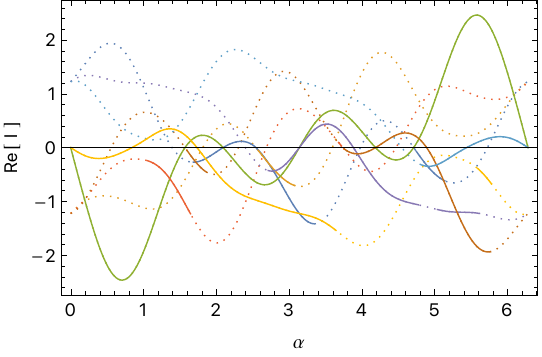}\hspace{1ex}
    \includegraphics[width=0.4\textwidth]{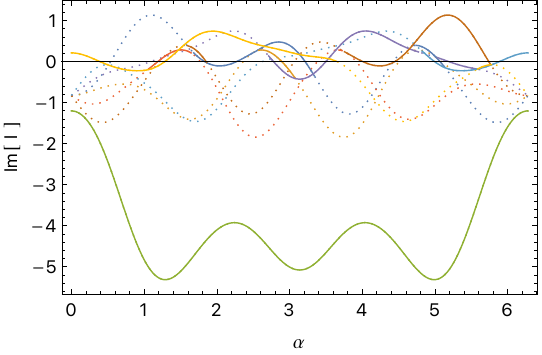}\hspace{1ex}
    \caption{The real part and the imaginary part of the exponent as functions of $\alpha$. The solid lines indicate $|n_\sigma|=1$ and the dotted lines indicate $|n_\sigma|=0$. The color scheme is the same as in Fig.~\ref{fig:airy_saddles}.}
    \label{fig:airy_exponent}
\end{figure}

Next, we compute the integral as a function of $\alpha$ using both the saddle-point approximation and direct numerical integration. We employ $\delta r = 0.01, 0.02, 0.03$ and $N = 200$ subintervals. The Newton's method is iterated for $100$ steps without line search, followed by $100$ steps with line search using $c_{\rm LS} = 1$. Saddle points are labeled to ensure continuity with respect to the parameter $\alpha$. The trajectories of the saddle points as $\alpha$ varies are depicted in Fig.~\ref{fig:airy_saddles}, where distinct saddle points are indicated by different colors. In the same color scheme, Fig.~\ref{fig:airy_exponent} presents the real and imaginary parts of the exponent evaluated at the saddle points as functions of $\alpha$.

For the direct numerical integration, we deform the contour from $\mathcal Y$ to $\tilde{\mathcal Y}$, defined as
\begin{equation}
    \tilde{\mathcal Y} = \{ z \in \mathbb{C}^3 \mid z_i = e^{i\theta\,\mathrm{sign}(y_i)} y_i,~y_i \in \mathbb{R} \},
\end{equation}
with $\theta = 0.1\pi$. The integral is evaluated using the Simpson rule with $250$, $300$, $350$, and $400$ intervals for each coordinate, and the integration range is systematically increased as $|y_i|/z_{\max} < 2.8, 2.9, 3.0, 3.1$, where $z_{\max}$ denotes the maximum modulus among the saddle points.

\begin{figure}[t]
    \centering
    \includegraphics[width=0.3\textwidth]{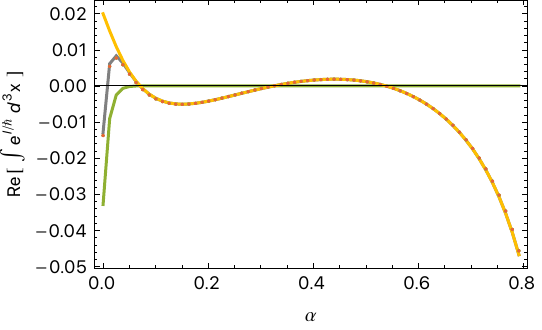}\hspace{1ex}
    \includegraphics[width=0.3\textwidth]{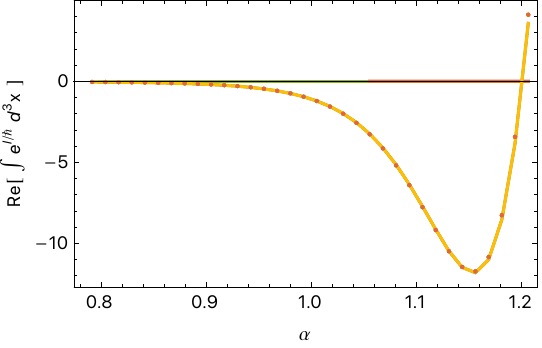}\hspace{1ex}
    \includegraphics[width=0.3\textwidth]{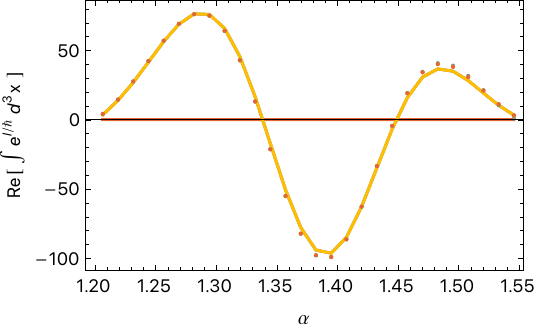}
    \includegraphics[width=0.3\textwidth]{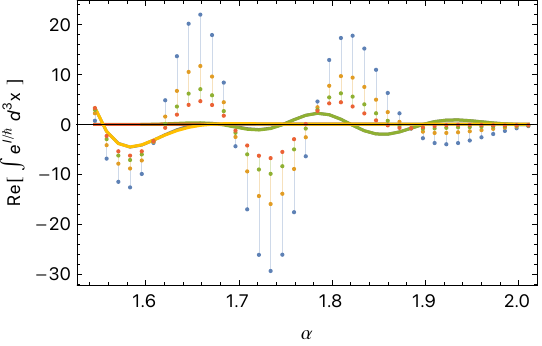}\hspace{1ex}
    \includegraphics[width=0.3\textwidth]{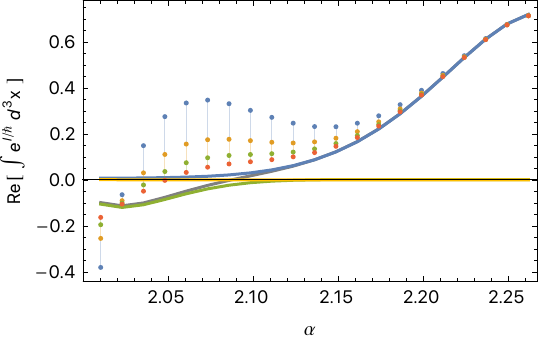}\hspace{1ex}
    \includegraphics[width=0.3\textwidth]{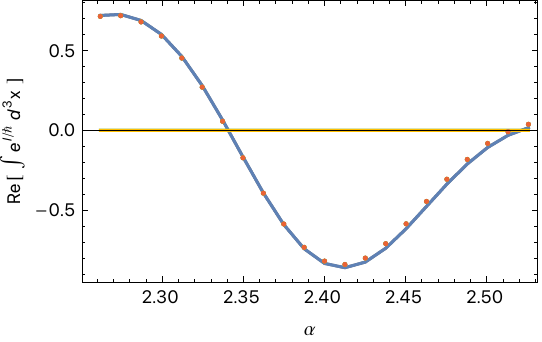}
    \includegraphics[width=0.3\textwidth]{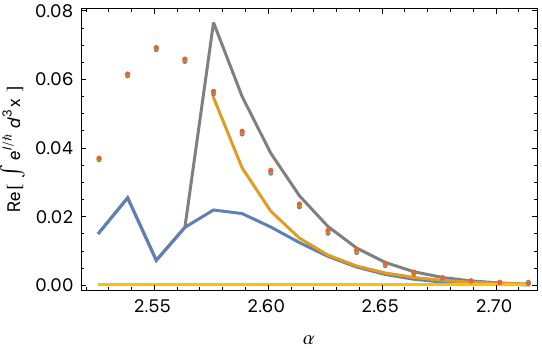}\hspace{1ex}
    \includegraphics[width=0.3\textwidth]{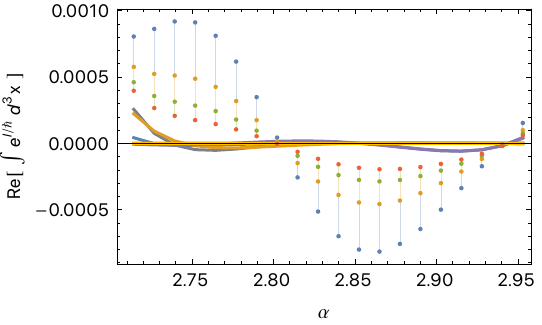}\hspace{1ex}
    \includegraphics[width=0.3\textwidth]{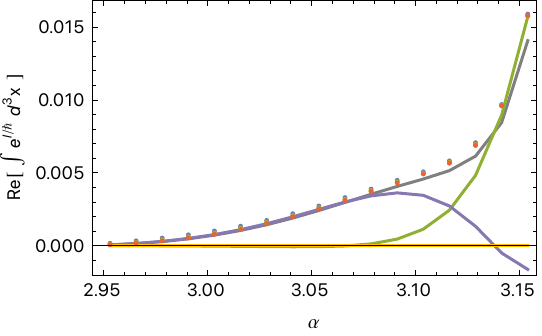}
    \includegraphics[width=0.3\textwidth]{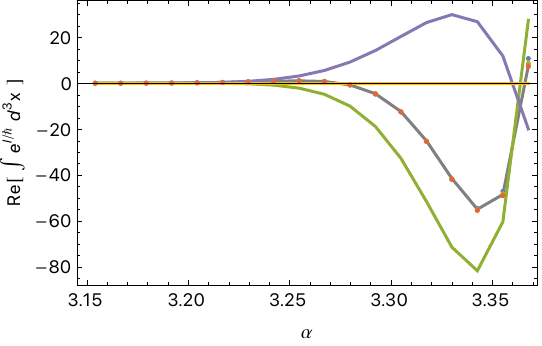}\hspace{1ex}
    \includegraphics[width=0.3\textwidth]{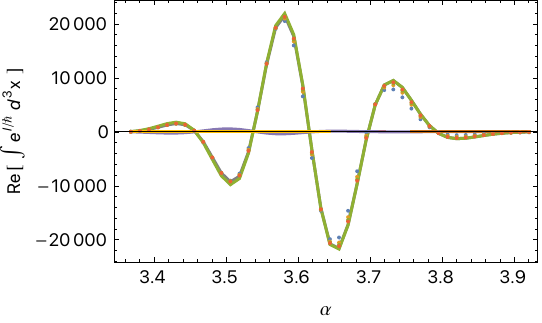}\hspace{1ex}
    \includegraphics[width=0.3\textwidth]{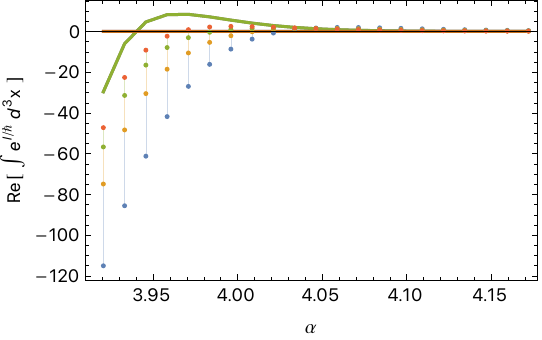}
    \includegraphics[width=0.3\textwidth]{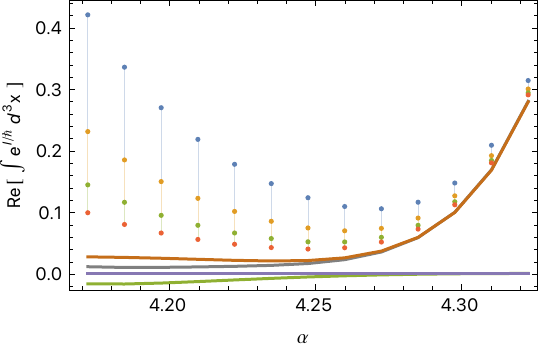}\hspace{1ex}
    \includegraphics[width=0.3\textwidth]{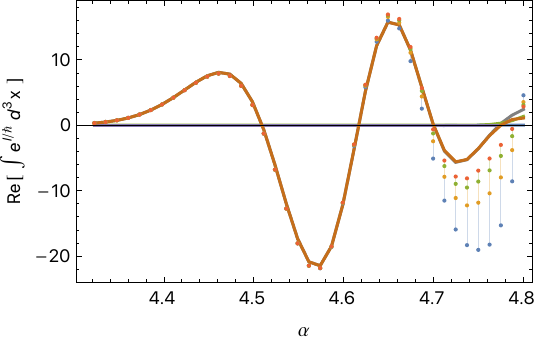}\hspace{1ex}
    \includegraphics[width=0.3\textwidth]{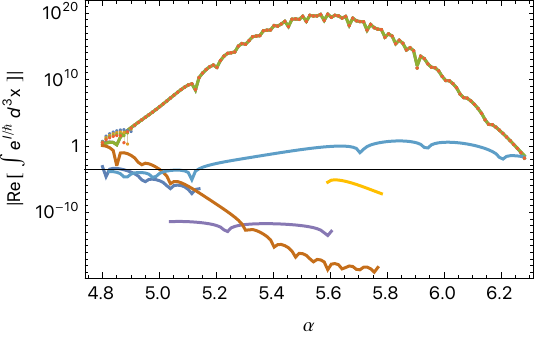}
    \caption{Values obtained from the numerical integration and from the saddle point approximation. The colored lines correspond to the contributions from the saddle points, where the color scheme is the same as in Fig.~\ref{fig:airy_saddles}. The gray line is the sum of these contributions. The blue, orange, green, and red points indicate the direct numerical computations for the different choices of lattice points and integration ranges, and the precision increases in this order.}
    \label{fig:airy_saddlepointapprox}
\end{figure}

We determine the intersection numbers, including their signs, and compute the saddle-point approximation. The results are summarized in Fig.~\ref{fig:airy_saddlepointapprox}, where the colored lines represent the contributions from individual saddle points, and the gray line shows their sum. The blue, orange, green, and red points correspond to the direct numerical computations for the different choices of lattice points and integration ranges.
Notice that when a saddle point is on the real plane, it gives $|n_{\sigma}|=1$ by definition.

We see that there are several regions where the numerical integration is hard. For example, around $\alpha=4.2$, the integral value is almost zero due to the cancellation of the oscillatory integrand. In this region, the numerical integration becomes unstable, and the results depend on the choice of lattice points and integration ranges.
However, the saddle-point approximation remains stable and accurately captures the behavior of the integral.

We also see that around $\alpha=3.12$ and $3.35$, two saddle points are constructively and destructively interfering, respectively. The sum of these contributions accurately reproduces the numerical results, demonstrating that we successfully compute the correct signs of the intersection numbers.

Near $\alpha = 2.58$, several noteworthy phenomena occur. First, two saddle points giving dominant contributions (orange and blue) approach each other, leading to reduced accuracy in the saddle-point approximation due to significant overlap of their Gaussian tails. Second, the flow structure between these saddles becomes highly intricate, and the flow originating from the orange saddle gets closer to the blue saddle and exhibits pronounced instability. This is reflected in a singular value of $\mathtt{R_S}$, which arises from the $J_Z^{(k)}$ matrices with small $k$ that deviate substantially from the identity.
Third, we observe a Stokes phenomenon for the orange saddle. This arises from a qualitative alteration in the flow structure, driven by the shifting relative positions of the saddle points as the parameter $\alpha$ varies.
This behavior is consistent with the numerical result: as the contribution from the orange saddle increases for $\alpha < 2.56$, it should cease to contribute beyond this point. It is important to note that the apparent discontinuity resulting from the change in intersection number is an artifact of the leading-order saddle-point approximation and will be smoothed out upon inclusion of higher-order corrections.


\section{Detailed analysis on double-well potential}
We present a detailed analysis of the path integral with the double-well potential discussed in the main text.
This system is based on the continuum formulation described in Ref.~\cite{Tanizaki:2014xba},
\begin{equation}
    \mathcal I_\infty[z]=\frac{i}{\hbar}\int_0^T\odif{t}\left[\frac{1}{2}\left(\odv{z}{t}\right)^2-\frac{1}{2}(z^2-1)^2\right],
\end{equation}
subject to the boundary conditions $z(0)=z(T)=0$.
The determination of the complex saddle points can be more efficiently achieved by solving the energy conservation equation, 
\begin{equation}
    \left(\frac{dz}{dt}\right)^2+(z^2-1)^2=p^2,
\end{equation}
with some $p\in \mathbb{C}$, as opposed to solving the Euler–Lagrange equation, 
\begin{equation}
    \frac{d^2z}{dt^2}=-2z(z^2-1).
\end{equation}
The solutions are then given in terms of Jacobi elliptic functions as
\begin{equation}
    z(t)=\sqrt{\frac{p^2-1}{2p}}\jsd\left(\sqrt{2p}\,t,\sqrt{\frac{1+p}{2p}}\right),
\end{equation}
where $p$ has to be chosen such that it satisfies the boundary condition $z(T)=0$. Introducing the parametrization $p = 1/(2k^2 - 1)$, 
where $k$ is called the elliptic modulus, the problem of determining $p$ is thereby reduced to identifying the corresponding value of $k$. Let us define the set of solutions as
\begin{equation}
    \mathcal{S}:=\Big\{z:[0,T]\rightarrow \mathbb{C} \Big| \ \frac{d^2z}{dt^2}=-2z(z^2-1), z(0)=z(T)=0\Big\}
\end{equation}
and another set 
\begin{equation}
    \Sigma:=\Big\{[(n,m)]\in \mathbb{Z}^2/\sim \Big| \ \frac{n}{\gcd (n,m)}\cdot \frac{m}{\gcd (n,m)} \equiv 0 \mod 2 \Big\},
\end{equation}
where $\sim$ is the equivalence relation $(n,m)\sim (-n,-m)$ and $\gcd (n,m)$ is the greatest common divisor, which is formally taken to be $1$ if $nm=0$. In the following, we fix the sign by $n>0$ for the cases where $n\neq 0$. 
In Ref.~\cite{Tanizaki:2014xba}, it has been shown that there is a one-to-one correspondence between the sets $\mathcal{S}/\{\pm1\}$ and $\Sigma$, via the transcendental equation
\begin{equation}
    n\omega_1(k)+m\omega_3(k)=\frac{T}{2},
\end{equation}
with
\begin{align}
    \omega_1(k)&=\sqrt{\frac{2k^2-1}{2}}K(k),\\
    \omega_3(k)&=i\sqrt{\frac{2k^2-1}{2}}K(\sqrt{1-k^2}),
\end{align}
where $K(k)$ denotes the complete elliptic integral of the first kind.

To resolve degeneracies among saddle points labeled by $(n,m)$ and $(m,-n)$, which yield identical values of $\mathcal I(z)$, we introduce a Morsification term to the action:
\begin{equation}
    \Delta\mathcal I[z]=ic\int_0^T\odif{t}\left[1+\left(\frac{t}{T}\right)^2\right]x(t),
\end{equation}
with its discretized form given in Eq.~\eqref{eq:morsification}. In our computations, we set $c=0.001 + 0.001 i$, although any sufficiently small parameter may be chosen provided it does not qualitatively alter the flow structure. We also checked our results are stable if we decrease $|c|$.
The explicit $t$-dependence is necessary to break residual symmetries, such as $t \to T-t$, and thereby ensure all saddle points are generic.

In our analysis, the time interval is discretized into $L$ segments, with variables defined as $z_i = z(T(i+1)/(L+1))$ for $i = 0, \ldots, L-1$. The saddle points on the lattice are then determined numerically, using the exact continuum solutions as initial guesses.
The number of segments $L$ has to be chosen appropriately for each saddle point. There exists a minimum value $L_{\min}$ below which the discrete approximation fails to accurately capture the structure of the saddle point, or the corresponding saddle point may not exist at all. We determine $L_{\min}$ by confirming that the shape of the saddle point does not change significantly as $L$ increases. Conversely, for cases where an intersection is present, there is a maximum value $L_{\max}$ beyond which the condition number of the Jacobian matrix in Newton's method becomes excessively large, resulting in numerical instability and unreliable inversion. This is diagnosed by monitoring $||\mathtt{R_F}||_{\max}$; values approaching the limit of double precision $10^{15}$ indicate that the Jacobian is ill-conditioned and cannot be stably inverted. We observe that once $||\mathtt{R_F}||_{\max}$ exceeds $10$, it rapidly increases and reaches $10^{15}$. For non-intersecting cases, $||\mathtt{R_F}||_{\max}$ typically remains small, and thus there is no clear $L_{\max}$.

The limitations described above also constrain the determination of intersection numbers. For intersecting cases, we observe that the intersection number remains nonzero as $L$ increases up to $L_{\max}$. (For some saddles, $L=L_{\max}-1$ has a larger $||\mathtt{R_F}||_{\max}$ and Newton's method fails to converge.) Based on this, we infer that the intersection number is likely preserved in the continuum limit $L \to \infty$, although a rigorous justification is beyond the scope of this work. Nevertheless, this observation provides valuable insight into the behavior of intersection numbers in the continuum case.

Conversely, establishing that the intersection number is zero is more subtle. If $L_{\max}$ is smaller than $L_{\min}$, it may not be possible to identify a flow even if one exists. To address this, we estimate $L_{\max}$ using the saddle points with similar $(n,m)$ and verify that no solution is found for $L$ within the estimated range. It is worth noting that the dependence of Newton's method on the initial guess is minimal, and the optimization sequence typically converges rapidly even for $L=20$.
To ensure the robustness, we also explore multiple initial guesses for the Newton's method. Thus, provided that the Jacobian remains well-conditioned, it is unlikely to miss a solution due to unfavorable initial conditions.

To demonstrate the robustness and reliability of our intersection number determination, we first benchmark our method against saddle points with analytically known intersection numbers. Unless otherwise specified, we set $N=300$, $T=5$ and $\delta r=0.01$ throughout our analysis. The initial guess for the step size is fixed as $\delta s=0.01$, while those for $Z^{(k)}$ are the same as in the Airy-type example. A single Dormand-Prince step is used for $\Phi$.
We iterate Newton's steps by $100$ without line search and another $100$ with line search. For $L=20$, the computation time is about $15\rm s$ with a single thread with a Rust implementation. The computational cost grows almost linearly in $L$ in this regime and is dominated by the computation of $J_z^{(k)}$. Notice that it formally grows as $L^2$, but operations on $L$ vectors are heavily optimized by the compiler.

The sign convention of the intersection number is derived from the path integral measure,
\begin{equation}
    \int\mathcal D xe^{\frac{\mathcal I(x)}{\hbar}}\simeq\int\frac{1}{\sqrt{2\pi i\Delta t}}\prod_{j=0}^{L-1}\frac{\odif{x_j}}{\sqrt{2\pi i\Delta t}}e^{\frac{\mathcal I(x)}{\hbar}}\simeq\sum_\sigma \frac{n_\sigma A_\sigma}{(2\pi i\Delta t)^{\frac{L+1}{2}}} e^{\frac{\mathcal I(z_\sigma)}{\hbar}}.
\end{equation}
Assuming that the sign of the overall factor is preserved in the continuum limit $L\to\infty$, we adopt the convention of the orientation of $\mathcal J_\sigma$ such that $\Re(A_\sigma/(2\pi i\Delta t)^{\frac{L+1}{2}})>0$, which determines the sign of $n_\sigma$. We also confirmed that this sign convention is stable across different $L$ for complex saddles. However, for real saddles, the sign is not very stable against the change of $L$. This is probably caused by contamination of other real saddles due to the lattice effects.

\begin{figure}[t]
    \centering
    \includegraphics[width=0.4\textwidth]{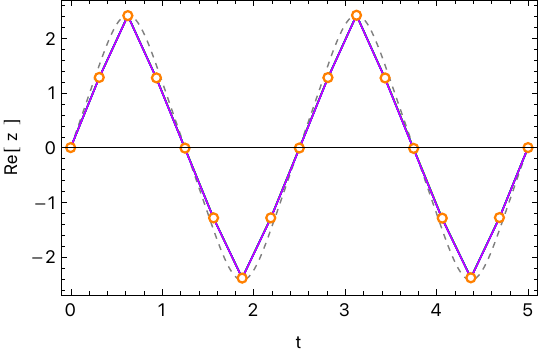}
    \includegraphics[width=0.4\textwidth]{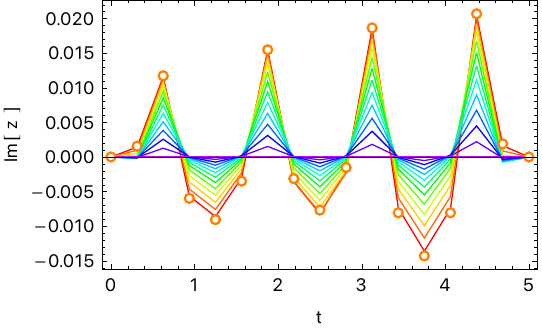}
    \caption{The same figure as in Fig.~\ref{fig:dw_flow}. We set $\delta r=0.003$, $c=0.1+0.1i$ and $(m,n)=(4,0)$.}
    \label{fig:dw_flow_sm_2}
\end{figure}

First, real solutions are obtained for $m=0$ or $n=0$, which always contribute to the integral. 
A small perturbation with the Morsification terms makes them complex, but they generically do not change the intersection numbers until crossing a Stokes line (except in the case where the real saddle point is exactly on the Stokes line). Thus, such saddle points give a false negative test for our method. We increase the magnitude of the Morsification parameter and see the intersection number is preserved. We confirm in Table \ref{table:dw_saddlesnm0} that $|n_\sigma|$ stays one for $c=0.1+0.1i$, where we take $\delta r=0.003$ to ensure that $\Im Z^{(0)}$ is close enough to $\Im Z_{\sigma}$. Further increasing its magnitude by a factor will cross a Stokes line and the intersection number changes. We show the obtained flow for $(n,m)=(4,0)$ in Fig.~\ref{fig:dw_flow_sm_2}. We do not show the sign of the intersection number since $L_{\max}-L_{\min}$ is small and also the sign depends on $L$ for some saddles. Notice that this does not mean the failure of the sign determination for each $L$: Since the real saddle points are almost like sine functions having no imaginary parts, they are easily contaminated by other saddles due to the lattice effects.

As a side note, a real saddle point $z_\sigma$, $\mathcal I_\infty(z_\sigma)$ is purely imaginary and hence $\Re[I(z)]>0$ for any $z\in \mathcal{K}_\sigma \setminus \{z_\sigma\}$ in the upward flowing cycle. This means that multiple intersections are not possible for non-degenerate real saddles. We numerically observed that all trials with different initial guesses converge to the same solution, consistently with the fact that there is no second intersection.

\begin{table}[t]
    \centering
    \begin{tabular}{c|c|c|c|c|c|c|c|c}
        $n$&$m$&$\mathcal I_\infty[z]$&$\mathcal I(z)$ for $L_{\max}$&$L_{\min}$&$L_{\max}$&$||\mathtt{R_F}||_{\max}$ for $L_{\max}$&$\mathcal R_{\rm tot}$ for $L_{\max}$&$|n_\sigma|$\\
        \hline
        $1$&$0$&$0. - 0.596 i$&$-0.345 - 0.590 i$&$3$&$8$&$1.5 \times  10^4$&$6.3\times 10^{-16}$&$1$\\
        $0$&$-1$&$0. - 0.596 i$&$0.336 - 1.385 i$&$3$&$7$&$8.9 \times 10^5$&$6.6\times 10^{-16}$&$1$\\
        $2$&$0$&$0.+2.457 i$&$0.102 + 1.843 i$&$7$&$10$&$3.0 \times 10^{6}$&$1.0\times 10^{-15}$&$1$\\
        $0$&$-2$&$0.+2.457 i$&$-0.098 + 2.044 i$&$7$&$10$&$1.0 \times 10^{9}$&$1.1\times 10^{-15}$&$1$\\
        $3$&$0$&$0. + 9.738 i$&$-0.241 + 8.191 i$&$10$&$11$&$2.6 \times 10^{3}$&$1.3\times 10^{-15}$&$1$\\
        $0$&$-3$&$0. + 9.738 i$&$0.243 + 7.707 i$&$10$&$11$&$1.6 \times 10^{3}$&$1.0\times 10^{-15}$&$1$\\
        $4$&$0$&$0. + 25.180 i$&$0.077 + 21.499 i$&$13$&$15$&$2.3 \times 10^{4}$&$2.3\times 10^{-15}$&$1$\\
        $0$&$-4$&$0. + 25.180 i$&$-0.078 + 21.654 i$&$13$&$15$&$3.0 \times 10^{4}$&$2.6\times 10^{-15}$&$1$
    \end{tabular}
    \caption{The intersection numbers for the first few real saddle points identified with $(n,m)$. Shown are also the values of $\mathcal I_\infty[z]$ obtained by integrating the continuous solution and $\mathcal I(z)$ the sum of the discretized and deformed action with Morsification parameter. We set $\delta r=0.003$ and  $c=0.1+0.1 i$.}
    \label{table:dw_saddlesnm0}
\end{table}

Second is the false positive test for our method. For $n\neq 0$ and $m \neq 0$, all the solutions are complex. For $n<m$ or $n>-m>0$, the saddle point fulfills $\Re I>0$ and hence also $\Re I>0$ along the upward flowing cycle. Therefore, the intersection number must be zero. We confirmed that our method stably gives $n_\sigma=0$ for this case, and the results are shown in Table \ref{table:dw_saddlesinters0}.

\begin{table}[t]
    \centering
    \begin{tabular}{c|c|c|c|c|c}
        $n$&$m$&$\mathcal I_\infty[z]$&$L_{\min}$&$n_\sigma$\\
        \hline
        $1$&$2$&$1.280 + 1.427 i$&$6$&$0$\\
        $2$&$-1$&$1.280 + 1.427 i$&$6$&$0$\\
        $1$&$4$&$14.926 + 19.727 i$&$10$&$0$\\
        $4$&$-1$&$14.926 + 19.727 i$&$10$&$0$\\
        $2$&$3$&$7.357 - 0.759 i$&$10$&$0$\\
        $3$&$-2$&$7.357 - 0.759 i$&$10$&$0$\\
        $2$&$4$&$23.946 + 4.198 i$&$12$&$0$\\
        $4$&$-2$&$23.946 + 4.198 i$&$12$&$0$\\
        $3$&$4$&$21.025 - 18.980 i$&$13$&$0$\\
        $4$&$-3$&$21.025 - 18.980 i$&$13$&$0$
    \end{tabular}
    \caption{The intersection numbers for the first few saddle points with $\Re I>0$. We set $\delta r=0.01$ and $c=0.001+0.001i$.}
    \label{table:dw_saddlesinters0}
\end{table}

\begin{figure}[t]
    \centering
    \includegraphics[width=0.4\textwidth]{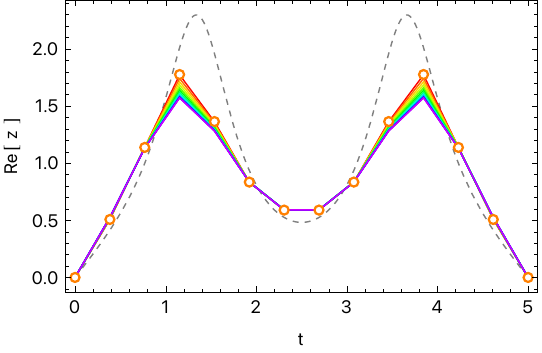}
    \includegraphics[width=0.4\textwidth]{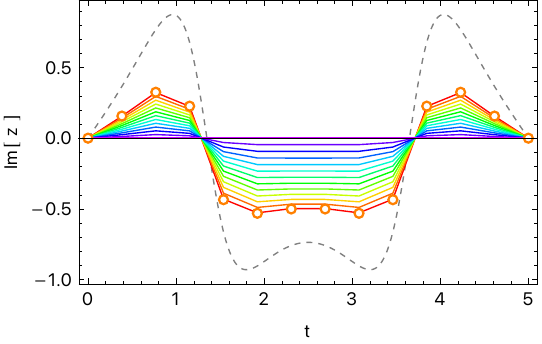}
    \includegraphics[width=0.4\textwidth]{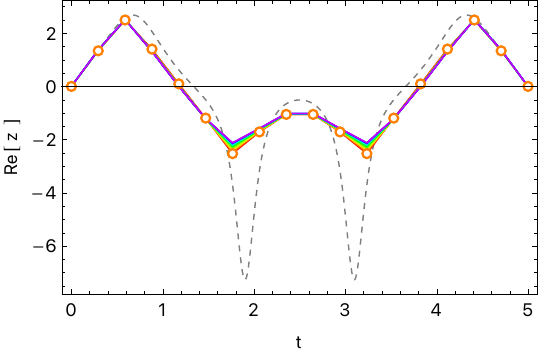}
    \includegraphics[width=0.4\textwidth]{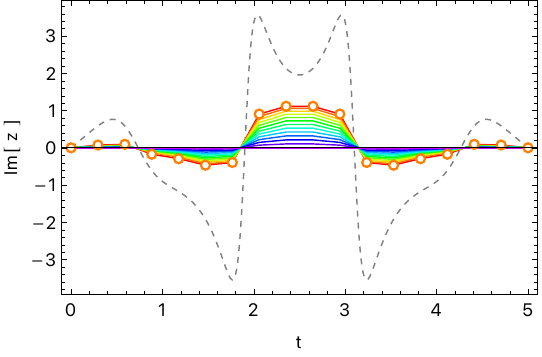}
    \caption{The same figure as in Fig.~\ref{fig:dw_flow}. The top panels are for $(m,n)=(2,1)$ and the bottom panels are for $(m,n)=(4,1)$. We set $\delta r=0.01$ and $c=0.001+0.001i$.}
    \label{fig:dw_flow_sm_1}
\end{figure}

We now turn to the nontrivial cases where $(n,m)$ or $(m,-n)$ with $n>m>0$, which yield $\Re\mathcal I<0$, resulting in intersection numbers that are not determined by general arguments. Using our method, we have computed the intersection numbers for the first several non-trivial saddle points; the results are summarized in Table~\ref{table:dw_saddlesdetailed}. For reference, a simplified version of this table is presented in the main text as Table~\ref{table:dw_saddles}. We show the obtained flows in Figs.~\ref{fig:dw_flow} and \ref{fig:dw_flow_sm_1}. The saddle of $(m,-n)$ has the same shape as $(n,m)$, but the overall sign is flipped. We find that the intersection numbers are stable in the range $L\in[L_{\min},L_{\max}]$, including their signs. This suggests that we are correctly tracking the same saddle point as $L\to\infty$: Unlike real saddles, complex saddles exhibit significantly different shapes for different $(n,m)$, making it easier to pin down the desired one.

\begin{table}[t]
    \centering
    \begin{tabular}{c|c|c|c|c|c|c|c|c}
        $n$&$m$&$\mathcal I_\infty[z]$&$\mathcal I(z)$ for $L_{\max}$&$L_{\min}$&$L_{\max}$&$||\mathtt{R_F}||_{\max}$ for $L_{\max}$&$\mathcal R_{\rm tot}$ for $L_{\max}$&$n_\sigma$\\
        \hline
        $2$&$1$&$-1.280 + 1.427 i$&$-0.775 + 1.271 i$&$6$&$12$&$4.4 \times 10^{13}$&$1.4\times 10^{-13}$&$+1$\\
        $1$&$-2$&$-1.280 + 1.427 i$&$-0.764 + 1.257 i$&$6$&$12$&$4.5 \times 10^{13}$&$9.7\times 10^{-13}$&$+1$\\
        $3$&$2$&$-7.357 - 0.759 i$&$-$&$10$&$-$&$-$&$-$&$0$\\
        $2$&$-3$&$-7.357 - 0.759 i$&$-$&$10$&$-$&$-$&$-$&$0$\\
        $4$&$1$&$-14.926 + 19.727i$&$-5.783 + 17.860  i$&$10$&$16$&$3.6 \times 10^{7}$&$5.8\times 10^{-15}$&$-1$\\
        $1$&$-4$&$-14.926 + 19.727i$&$-5.783 + 17.862 i$&$10$&$16$&$3.6 \times 10^{7}$&$5.4\times 10^{-15}$&$-1$\\
        $4$&$2$&$-23.946 + 4.198 i$&$-15.311 + 6.545 i$&$12$&$20$&$1.1 \times 10^{13}$&$1.6\times 10^{-13}$&$-1$\\
        $2$&$-4$&$-23.946 + 4.198 i$&$-15.314 + 6.549  i$&$12$&$20$&$1.1 \times 10^{13}$&$1.4\times 10^{-13}$&$-1$\\
         $4$&$3$&$-21.025 - 18.980 i$&$-$&$13$&$-$&$-$&$-$&$0$\\
        $3$&$-4$&$-21.025 - 18.980 i$&$-$&$13$&$-$&$-$&$-$&$0$
    \end{tabular}
    \caption{The intersection numbers for the first few non-trivial complex saddle points. We set $\delta r=0.01$ and $c=0.001+0.001i$.}
    \label{table:dw_saddlesdetailed}
\end{table}

\end{document}